\documentstyle[aaspp4,epsf]{article}

\begin{document}

%\rightline{EFI-TH-97}
 
\title{Interferometric Observation of Cosmic Microwave Background
Anisotropies}
\author{Martin White${}^1$, John E Carlstrom${}^2$, Mark Dragovan${}^2$,
William L Holzapfel${}^2$}
\affil{
${}^1$Dept of Astronomy and Dept of Physics,\\
University of Illinois at Urbana-Champaign, Urbana IL 61801\\
${}^2$Dept of Astronomy and Astrophysics,\\
University of Chicago, Chicago, IL 60637}
\authoremail{white@physics.uiuc.edu}

\begin{abstract}
\noindent
\rightskip=0pt
We present a formalism for analyzing interferometric observations of the
Cosmic Microwave Background (CMB) anisotropy and polarization.
The formalism is based upon the $\ell$-space expansion of the angular
power spectrum favored in recent years.
Explicit discussions of maximum likelihood analysis, power spectrum
reconstruction, parameter estimation, imaging and polarization are given.
As an example, several calculations for the Degree Angular Scale
Interferometer (DASI) and Cosmic Background Interferometer (CBI) experiments
are presented.
\end{abstract}


\rightskip=0pt
\section{Introduction}

The Cosmic Microwave Background (CMB) has become one of the premier
tools for understanding early universe astrophysics, classical cosmology
and the formation of large-scale structure.
It has a wealth of information about the origin and evolution of the
Universe encrypted in its signal.  The frequency spectrum is that of a
blackbody at 2.73K, confirming the prediction of the standard hot big
bang model (see Fixsen et al.~\cite{Spectrum}, Nordberg \& Smoot~\cite{NorSmo}
or the review of Smoot \& Scott~\cite{SmoSco}).
The small angular scale power spectrum of the temperature and polarization
anisotropies contain information on the cosmological parameters and the
structure that existed at decoupling
(see e.g.~Bennett, Turner \& White~\cite{BenTurWhi}).

The predictions for a wide range of cosmological models are now well
understood and theoretically secure (Hu et al.~1997; hereafter \cite{OTAMM}).
Experimentally, a flurry of results have been reported in the last five years.
The study of the microwave sky has moved beyond the ``detection phase'' and
into the ``imaging phase'', with the next generation of experiments planning
to provide detailed information about the shape of the angular power spectrum
of the temperature anisotropies over a wide range of angular scales.

The advent of low-noise, broadband, millimeter-wave amplifiers
(Popieszalski~\cite{Pop}) has made interferometry a particularly attractive
technique for detecting and imaging low contrast emission, such as anisotropy
in the CMB.
An interferometer directly measures the Fourier transform of the intensity
distribution on the sky.
By inverting the interferometer output, images of the sky are obtained which
include angular scales determined by the size and spacing of the individual
array elements.
In this paper we discuss a formalism for interpreting CMB anisotropies as
measured by interferometers and examine what two upcoming experiments,
the Degree Angular Scale Interferometer (DASI) and Cosmic Background
Interferometer (CBI), may teach us about cosmology. 

Several previous papers have dealt with the analysis of CMB data from
interferometers
(Martin \& Partridge~\cite{MarPar}; Subrahmanyan et al.~\cite{SESS};
Hobson, Lasenby \& Jones~\cite{HobLasJon}; Hobson \& Magueijo~\cite{HobMag}).
In this paper we extend the work to make explicit contact with the multipole
space ($\ell$-space) methods now commonly adopted in analyzing single-dish
switching experiments, including power spectrum estimation, parameter
extraction, imaging, polarization and mosaicing.
We also give details of the upcoming DASI experiment.

The outline of the paper is as follows.
We begin with a discussion of instruments, past present and future, in
\S\ref{sec:instruments}.  Foregrounds and point sources are discussed in
\S\ref{sec:foregrounds}.  The theoretical formalism for analyzing temperature
anisotropies is outlined in \S\ref{sec:formalism} while applications including
maximum likelihood estimation of parameters and power spectrum reconstruction
are treated in \S\ref{sec:comparetheory}.
Increasing the sky coverage, and hence the resolution of the instrument in
$\ell$-space is introduced in \S\ref{sec:mosaic}.
Making images of the microwave sky is addressed in \S\ref{sec:mapping}.
Polarization is treated in \S\ref{sec:polarization}.
Finally, \S\ref{sec:discussion} contains our summary and discussion.

\section{Instruments} \label{sec:instruments}

The use of interferometers to study fluctuations in the CMB goes back over a
decade (see Table~\ref{tab:int}).
Early work using the VLA and ATCA concentrated on small angular scales,
reporting a series of upper limits.
Recently the CAT (O'Sullivan et al.~\cite{OSul}) has reported a detection of
anisotropy on sub-degree scales, at low frequencies ($15\;$GHz).
Several groups are now planning to build interferometers which operate at
higher frequencies and over a larger range of angular scales, with
sensitivities which should enable them to map in detail the CMB anisotropy
spectrum from $\ell\sim10^2$ to $\ell\sim10^3$.

\begin{table}[htb]
\begin{center}
\begin{tabular}{llccccc}
Name & Location & $N_{\rm dish}$ & Freq. (GHz) & Bandwidth (GHz) & Primary beam
 & $\ell$ \\ \hline
OVRO\tablenotemark{a}  & United States  &  6 & 30     & 2.0 & $4^{\prime}$ &
 6750 \\
VLA\tablenotemark{b}   & United States  & 27 & 8      & 0.2 & $5^{\prime}$ &
 6000 \\
Ryle\tablenotemark{c}  & England        &  8 & 15     & 0.4 & $6^{\prime}$ &
 4500 \\
BIMA\tablenotemark{d}  & United States  & 10 & 30     & 0.8 & $6^{\prime}$ &
 4300 \\
ATCA\tablenotemark{e}  & Australia      &  6 & 9      & 0.1 & $8^{\prime}$ &
 3400 \\
T-W\tablenotemark{f}   & United States  &  2 & 43     & --- & $2^{\circ}$  &
 20-100 \\
CAT\tablenotemark{g}   & England        &  3 & 13--17 & 0.5  & $2^{\circ}$ &
 339--722 \\ \hline
VSA\tablenotemark{h}   & Canary Islands & 15 & 26--36 & 2.0  & $4^{\circ}$ & 
130--1800 \\
DASI\tablenotemark{i}  & South Pole     & 13 & 26--36 & 10.0 & $3^{\circ}$ & 
125--700 \\
CBI\tablenotemark{j}   & Chile          & 13 & 26--36 & 10.0 & $44'$ & 
630--3500 \\ \hline
\end{tabular}
\end{center}
\tablenotetext{a}{Carlstrom, Joy \& Grego 1996, ApJ, 461, L59}
\tablenotetext{b}{Fomalont EB., et al., 1984, ApJ, 277, L23;
Knoke JE., et al., 1984, ApJ, 284, 479;
Martin HM. \& Patridge RB., 1988, ApJ, 324, 794;
Fomalont EB., et al., 1988, Astron J., 96, 1887;
Hogan CJ. \& Partridge RB., 1989, ApJ, 341, L29;
Fomalont EB., et al., 1993, ApJ, 404, 8.;
Partridge RB, et al., 1997, ApJ, 483, 38}
\tablenotetext{c}{Jones, ME., 1997, PPEUC proceedings, Cambridge, April 7-11}
\tablenotetext{d}{Cooray, AR. et al. 1997, AAS, 191 1906}
\tablenotetext{e}{Subrahmanyan R., Ekers RD., Sinclair M., Silk J., 1993,
MNRAS, 263, 416; Subrahmanyan R., Kesteven MJ., Ekers RD., Silk J., 1998,
MNRAS, in press [astro-ph/9805245]}
\tablenotetext{f}{Timbie P.T., Wilkinson D.T., 1990, ApJ, 353, 140}
\tablenotetext{g}{O'Sullivan C., et al., 1995, MNRAS, 274, 861}
\tablenotetext{h}{see Jones, ME., 1996, in Proceedings of
the XVIth Moriond Meeting, ed.~F.~Bouchet et al., p. 161}
\tablenotetext{i}{see Halverson et al., 1998,  ASP Conf. Proc.,
``Astrophysics from Antarctica'', ed. G.Novak \& R.Landsberg, in press}
\tablenotetext{j}{http://astro.caltech.edu/$\sim$tjp/CBI/}
\caption{Current experiments to measure CMB temperature anisotropies with
interferometers.  There are published upper limits from VLA, Ryle and ATCA.
The CAT has published a detection while VSA, DASI and CBI are expected
to begin operations around 1999-2000.
The location, number of dishes/horns, frequency and (approximate) coverage
in $\ell$ space are listed.}
\label{tab:int}
\end{table}

An interferometric system offers several desirable features:
1) It {\it directly\/} measures the power spectrum of the
sky, in contrast to the differential or total power measurements.
Images of the sky can then be created by aperture synthesis.
2) Interferometers are intrinsically stable since only correlated signals
are detected; difficult systematic problems that are inherent in total power
and differential measurements are absent in a well designed interferometer.
This considerably reduces signals due to ground pickup and near field
atmospheric emission.
3) They can be designed for continuous coverage of the CMB power spectrum
with angular spectral resolution determined by the number of fields imaged.

Motivated by these attributes, several groups are developing new
interferometers targeted at measuring the anisotropy in the CMB on
sub-degree angular scales.
Three instruments are currently under construction: the Very Small Array (VSA)
in Cambridge, the Degree Angular Scale Interferometer (DASI) and the
Cosmic microwave Background Interferometer (CBI).
DASI and CBI are parallel projects based at Chicago and Caltech.
In this paper we will concentrate on the DASI instrument, though the formalism
is completely general and applies equally to any interferometer.

\begin{figure}
\begin{center}
\leavevmode
\epsfysize=6cm \epsfbox{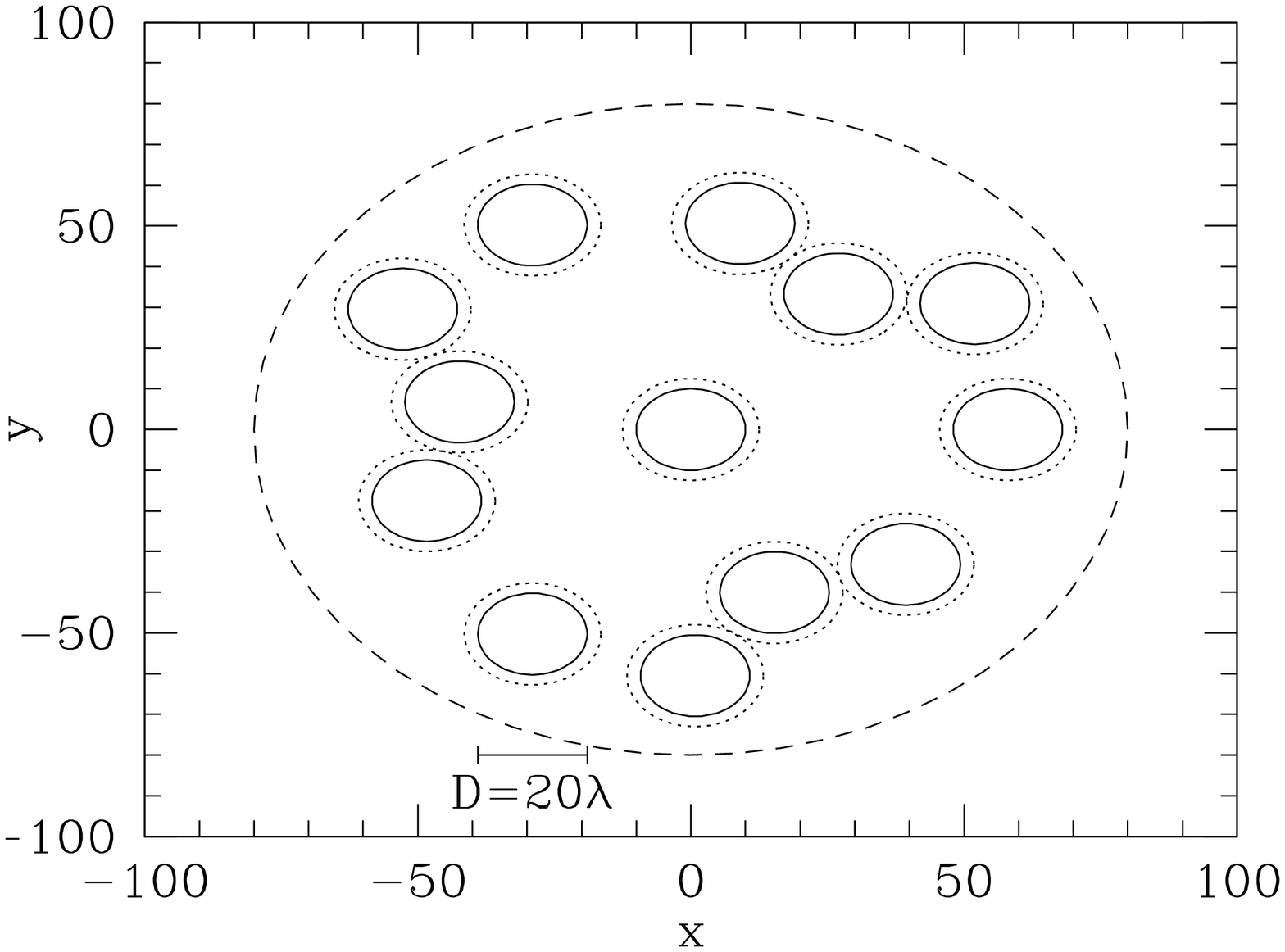}
\epsfysize=6cm \epsfbox{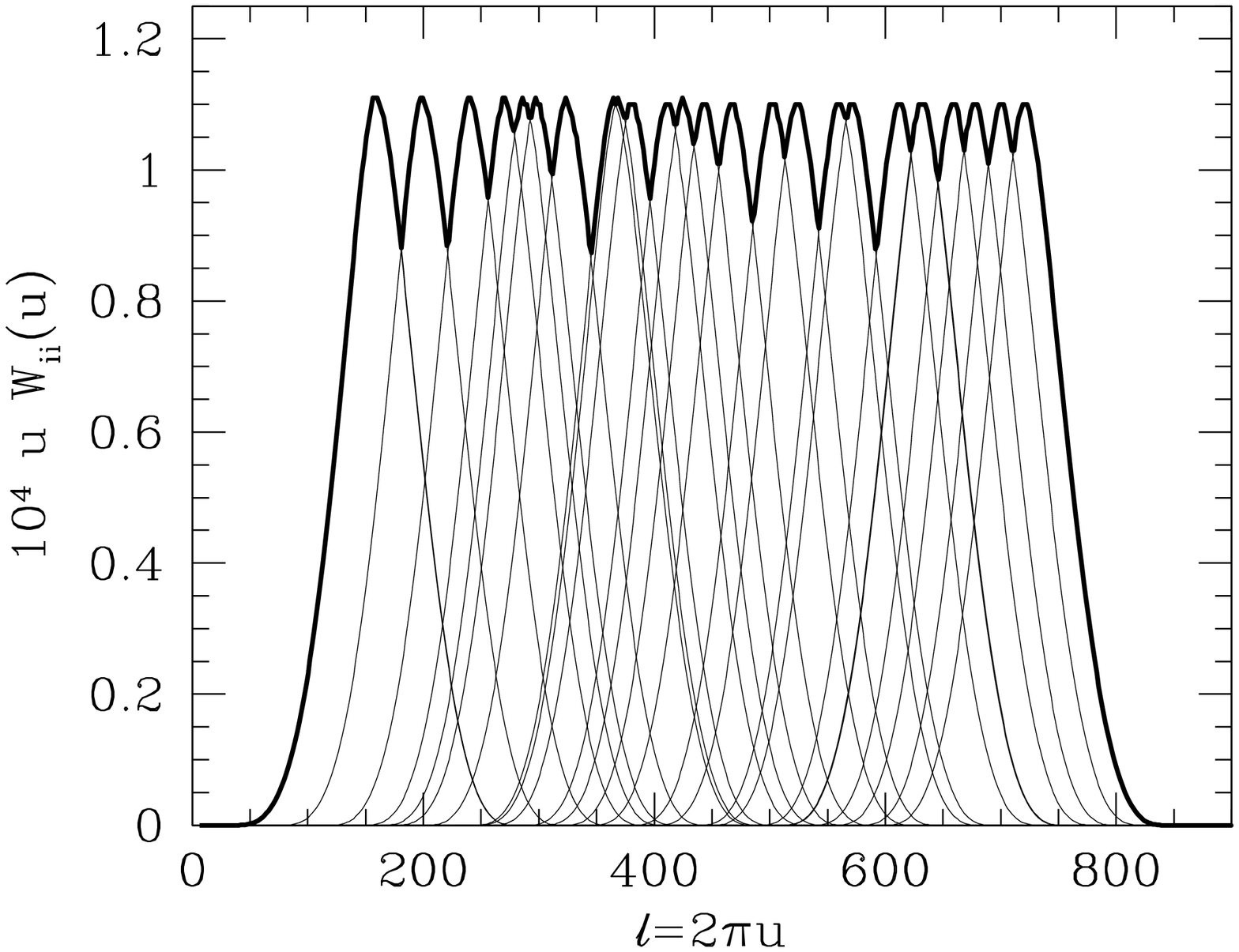}
\end{center}
\caption{(a) The physical configuration of the DASI receivers.  The axes
are graduated in wavelengths, physical separations are obtained by
multiplying by $\lambda=1\;$cm.
Each dish has an effective optical diameter $D=20\lambda$ (solid lines).
The physical size of the dishes is 25cm (dotted lines) and the base plate
for DASI is 1.6m in diameter (dashed line).
Correlations are measured between all pairs of horns, so the instrument
samples 78 baselines.  Note the 3-fold symmetry of the array, which means
that 26 different $u$'s are sampled.
(b) The sensitivity of DASI vs $\ell=2\pi u$ for the configuration in (a).
The window functions are shown for each of the 26 different $u$'s sampled,
and the bold line traces the outline of these 26 window functions.}
\label{fig:visibilities}
\end{figure}

DASI is an interferometer designed to measure anisotropies in the CMB
over a large range of scales with high sensitivity\footnote{More
information on DASI can be found at {\tt http://astro.uchicago.edu/dasi}.}.
The array consists of 13 closely packed elements, each of $20\;$cm diameter,
in a configuration which fills roughly half of the aperture area with a
3-fold symmetry (see Fig.~\ref{fig:visibilities}).
Each element of the array is a wide-angle corrugated horn with a collimating
lens.  DASI uses cooled HEMT amplifiers running between 26-$36\;$GHz with a
noise temperature of $<15$K.  The signal is filtered into ten $1\;$GHz
channels.  DASI will operate at the South Pole.

\section{Foregrounds} \label{sec:foregrounds}

In order to estimate the contribution of primordial fluctuations to the
observed signal, it is necessary to estimate the contribution from foreground
contaminants.
Some reviews of the situation with regards astrophysical foregrounds can be
found in Brandt et al.~(\cite{Brandtetal}),
Toffolatti et al.~(\cite{TofetalA,TofetalB}) and
Tegmark \& Efstathiou~(\cite{TegEfs}).
We refer the reader to these papers for more details, and lists of references.
Below we summarize some of the more important points for the DASI experiment.

\subsection{Atmosphere}

The far field\footnote{The farfield distance is $2 B^2/\lambda$ where
$B$ is the longest baseline of the interferometer, $1.1$m for DASI, and
$\lambda$ is the wavelength.} of a very compact array is actually quite near
the instrument; for DASI this distance is only a few hundred meters.
Thus everything beyond this distance, including the atmosphere, will be
imaged by the interferometer.
Taking a standard model for the static brightness of the atmosphere with a
temperature $T$ and zenith opacity $n$, the brightness varies with zenith
distance $\theta$ as
\begin{equation}
  T\left( 1-\exp\left[{-n\over\cos(\theta)}\right] \right)
\quad .
\end{equation}
Its effect thus appears as a constant term plus a slope and a very slight
curvature.
Since the interferometer rejects low spatial frequencies, this atmospheric
contribution is negligible in the final image.

The dynamic atmosphere causes a fluctuating brightness we must look through,
and if it is in the far field it will be correlated and appear as excess
noise (Church~\cite{Church}).
The Python V experiment observed from the South Pole during the austral summer
of 1997 at a frequency of $45\;$GHz and covered angular scales comparable to
those of DASI.
{}From the level of atmospheric noise, we estimate that the atmosphere
will contribute only about 10\% to the total system noise
(Lay et al., 1998, in preparation).

\subsection{Galactic Foregrounds}

\subsubsection{Synchrotron}

Due to its low operating frequency, the main foreground which DASI will
need to contend with is Galactic synchrotron radiation, produced by electrons
spiraling in the galactic magnetic field.
The specific intensity\footnote{To obtain the spectral indices in terms of
antenna temperature $T_A\equiv c^2 I_\nu/(2k\nu^2)$, or
$\delta T_A = x^2 e^x/(e^x-1)^2 \delta T$, one subtracts 2.}
of synchrotron emission roughly follows a power-law in frequency
$I_\nu^{\rm sync} \propto \nu^\beta$
with spatially varying index and amplitude
(Lawson et al.~\cite{Lawetal}, Banday \& Wolfendale~\cite{BanWol},
Platania et al.~\cite{Plaetal}).
The index, $\beta_{\rm sync}$, varies from -0.1 to -1.3 with a mean of -0.8.
There is some evidence that the index steepens at higher frequencies.
The angular power spectrum $C_\ell\sim\ell^{-3}$ for $\ell<10^2$ (see
Eq.~(\ref{eqn:cordef}) for a definition of $C_\ell$).

\subsubsection{Free-Free}

Free-free emission, also known as bremsstrahlung, is due to scattering of
unbound particles, typically electrons off nuclei e.g.~$ep\to ep\gamma$.
The spectral index $\beta_{\rm ff}$ depends on the temperature and density
of the charged particles, but is in the range -0.13 to -0.16 for typical
electron density and temperature values for the ISM
(Bennett et al.~\cite{Benetal}).
At high galactic latitudes, free-free emission is expected to dominate over
synchrotron emission at around $40\;$GHz (Bennett et al.~\cite{Benetal}).
No direct maps of free-free emission exist, though there is a possible
correlation between free-free emission and H$\alpha$ emission
(Bennett et al.~\cite{Benetal}).
Since the H$\alpha$ maps contain striping the significance of the correlation
is not easy to assess.
If there is a strong correlation the free-free spectrum can be predicted from
H$\alpha$ measurements at galactic latitude $\sim20^\circ$
(Reynolds~\cite{Reynolds}) plus fundamental physics to determine
$\beta_{\rm ff}$.
There is evidence however
(Kogut et al.~\cite{KogHBBGSW}, Leitch et al.~\cite{LRPM},
de Oliveira-Costa et al.~\cite{Angelica}, Kogut~\cite{Kogut})
that free-free emission may be correlated with dust emission near the NGP,
and that this ``hot'' ($10^5-10^6$K) component may not emit H$\alpha$
(for an alternative explanation of the correlation in terms of spinning
dust grains, see Draine \& Lazarian~\cite{DraLaz}).
A correlation between free-free emission and dust would
imply $C_\ell\sim\ell^{-2.5}$ to $\ell^{-3}$
(Schlegel et al.~\cite{SchFinDav}, Wright~\cite{Wri}).

\subsection{Extragalactic Foregrounds}

The fluctuations from extragalactic sources have been modeled by
Toffolatti et al.~(\cite{TofetalA,TofetalB}) and
Franceschini et al.~(\cite{Franetal}).
The source models are robust below $100\;$GHz though uncertain to almost
an order of magnitude well above $100\;$GHz.

The angular dependence of uncorrelated point sources is of course that of
white noise: $C_\ell\sim\ell^0$.
There is some evidence that radio sources are correlated
(Peacock \& Nicholson~\cite{PeaNic}), but the non-Poisson contribution to the
anisotropy is always smaller than the Poisson contribution
(Toffolatti et al.~\cite{TofetalA,TofetalB}).  Providing the sources exist
over a large range of distances from us, any correlation in the sources at
small scales is significantly diluted by projection.

\begin{figure}
\begin{center}
\leavevmode
\epsfysize=6cm \epsfbox{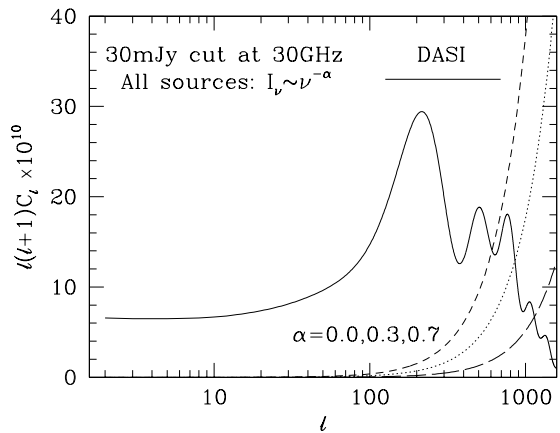}
\end{center}
\caption{The angular power spectrum of the standard CDM model (solid) in
dimensionless units.  The horizontal line represents the range of angular
scales probed by DASI.  The dashed and dotted lines are the angular power
spectrum of fluctuations from unresolved point sources.  These have been
modeled by taking the luminosity function of VLA FIRST
(Becker et al.~\protect\cite{VLAFIRST}) radio sources at $1.5\;$GHz as provided
by Tegmark \& Efstathiou (\protect\cite{TegEfs}; Eq.~43) and extrapolating
{\it all\/} sources to $30\;$GHz using $I_\nu\sim\nu^{-\alpha}$ with
$\alpha=0$, 0.3, 0.7.
All sources brighter than 30mJy at $30\;$GHz have been removed.}
\label{fig:pointsources}
\end{figure}

We show in Fig.~\ref{fig:pointsources} the angular power spectrum,
$\ell(\ell+1)C_\ell\sim\ell^2$, associated with point sources at $30\;$GHz,
assuming that we subtract all sources brighter than $30$mJy.
We have taken the luminosity function of VLA FIRST
(Becker et al.~\cite{VLAFIRST}) radio sources at $1.5\;$GHz from
Tegmark \& Efstathiou (\cite{TegEfs}; Eq.~43) and extrapolated {\it all\/}
the sources assuming $I_\nu\sim\nu^{-\alpha}$.  For $\alpha>0$ it is below
the expected cosmological signal for the range of scales probed by DASI.
Point source subtraction for DASI will be facilitated by using the ATCA
to map the DASI observing region at 16-26 \& $43\;$GHz
(the DASI observing region also overlaps with regions observed by Python
and planned observations by Boomerang and Beast).

\subsection{Foreground Subtraction}

We will for definiteness here consider the DASI instrument, though our
general conclusions will hold for the other planned instruments with similar
frequency coverage.
In the absence of external information about foreground emission we can use
the $10\;$GHz bandwidth of DASI to marginalize over the unknown amplitude of a
foreground component.  Since we are working at low frequency the dominant
foregrounds are synchrotron and free-free emission, which have similar
spectral indices.  Thus one is led to consider fitting out a single component.
Using the formalism of Dodelson~(\cite{Dod}) and 5 frequency channels centered
at 27, 29, 31, 33 and $35\;$GHz we find that the error bars on the CMB
component are increased by a factor of 4.2 (5.1) if we project out a
synchrotron (free-free) foreground with (temperature) spectral index -2.7
(-2.2).
Using $10\times 1\;$GHz channels doesn't change this number significantly.
With an improved foreground extraction method it may be possible to reduce
this by roughly a factor of 2 (Tegmark~\cite{TegMap}, White~\cite{WhiFor}).

The increase in the error bar is due to the small operating band of DASI,
and indicates that we would like to use external information (foreground maps)
when interpreting the DASI observations.
As an example, assumptions about the spatial properties of the foregrounds
(e.g.~their $\ell$-space power spectra) can improve the separation of
foreground and signal.
The DASI will originally operate in a region known to be low in foregrounds,
but the need for a large sky coverage will eventually require operation in
regions with significant foregrounds.
Due to the dependence on the region of sky surveyed, we will focus on the
instrument characteristics from now on, and not include the increase in the
error bars expected from foreground subtraction.

\section{Formalism} \label{sec:formalism}

\subsection{The Visibility}

An interferometer measures $\langle E_1 E_2^{*}\rangle$ where $E_1$ and
$E_2$ are the electric field vectors measured by two telescopes pointing
to the same position on the sky and $\langle\cdots\rangle$ represents an
average over a time long compared with the period of the wave.
Assuming for now a monochromatic source of radiation and working in the
Fraunhofer limit, the average of the product of electric fields is the
intensity times a phase factor.  For each point source, the phase factor is
the exponential of ($i$ times) the geometric path difference between the
source and the two telescopes, in units of the wavelength.
Taking the integral over the source/emitter plane gives the Fourier Transform
of the observed intensity
(i.e.~the sky intensity multiplied by the instrument beam).

The fundamental observable for the interferometer is thus a ``visibility'',
which is the Fourier Transform (FT) of the sky intensity multiplied by the
primary beam or aperture function
(Tompson, Moran \& Swenson~\cite{TomMorSwe}):
\begin{equation}
V(\vec{u}) \propto \int d\hat{x}\ A(\hat{x}) \Delta T(\hat{x})
	e^{2\pi i\vec{u}\cdot\hat{x}}
\label{eqn:visdef}
\end{equation}
where $\Delta T$ is the temperature (fluctuation) on the sky, $\hat{x}$ is
a unit 3-vector and $\vec{u}$ is the conjugate variable, with dimensions of
inverse angle measured in wavelengths.
$A(\hat{x})$ is the ``primary'' beam and is typically normalized to unity at
peak, which is $\sqrt{2\pi}\sigma$ larger than the usual normalization of a
gaussian beam. (By requiring $A(0)=1$ we ensure that the area of the aperture
in the $\vec{u}$ plane is unity.)
The spacing of the horns and the position of the beam on the sky determine
which value of $\vec{u}$ will be measured by a pair of antennae in any one
integration.
The size of the primary beam determines the amount of sky that is viewed,
and hence the size of the ``map'', while the maximum spacing determines the
resolution.

Typically the field of view of the interferometer is small, so in what
follows we will make a small angle approximation and treat the sky as flat
(see also the Appendix).
This is a very good approximation for the upcoming experiments (CBI, DASI, VSA)
and leads to significant simplification in the formalism.
If the primary beam $A(\hat{x})$, or more generally the area of sky surveyed,
is well localized the integral is only over a very small range of $\hat{x}$.
Denoting the center of the beam by $\hat{x}_0$ we can write
$\hat{x}=\hat{x}_0+{\bf x}_{\perp}$ with $x_\perp\ll x_0=1$ and
${\bf x}_\perp\cdot\hat{x}_0=0$, then ${\bf x}_\perp$ is a 2D vector lying
in the plane of the sky.
We will denote 2D vectors by boldface type, and vectors in 3D or other spaces
by arrows.  A vector name which is neither boldface nor arrowed indicates the
length of that vector, e.g.~$x_\perp=|{\bf x}_\perp|$.

Finally we should remark on one subtlety in the statistical analysis of a
``random'' component such as the CMB fluctuations.
It is common to assume that the temperature fluctuations in the CMB are
realizations of a random field, so that $T({\bf x})$ is a real random field
on $R^2$ (see below).
Since $V(\vec{u})$ is the Fourier Transform of $T({\bf x})$, it is also a
random field, however it is complex.
Since $T$ is real, it follows that $V^{*}(\vec{u})=V(-\vec{u})$ and thus
$V(\vec{u})$ is a complex random field with independent degrees of freedom
only over the half-plane (i.e.~twice as many as $T$ over half the area).
This restriction to the half-plane will be important when it comes to
parameter estimation.
An alternate formulation of the problem, which turns out to be equivalent,
is to define a ``real'' visibility in terms of cosine (sine) transforms
whenever e.g.~$u_x>0$ ($u_x<0$).  We will not give this parallel development
here, as it is exactly equivalent to the complex case we will discuss.

\subsection{The Sky Power Spectrum}

To proceed, notice that in the small field of view approximation the
visibility is the convolution of the FT of the sky intensity (temperature)
and the FT of the primary beam.
Thus if we knew the power spectrum of the sky, we could find the power
spectrum of the visibilities by convolution with the Fourier transform of
the primary beam (see below).
In this section we will concentrate on the sky power spectrum and neglect
the effect of the primary beam, and so we set $A=1$ for now.

We usually assume that our theory has no preferred direction, i.e.~it is
rotationally invariant.
In the flat sky approximation this rotational invariance becomes a
translational invariance on the plane.
This means that the (double) FT of the sky correlation function becomes
diagonal in ${\bf u}$, i.e.~``conserves momentum''.
(We discuss the flat sky approximation further in the Appendix.)
We shall call the diagonal part the sky power spectrum $S({\bf u})=S(u)$,
not to be confused with a flux.

The ability to perform Fourier analysis on the ``flat'' sky and the
replacement of rotational by translational invariance is the principle
advantage of the flat sky approximation.
These advantages are {\it only\/} obtained in the small-angle limit,
regardless of how one chooses to map angle into $u$, so the reader should
beware of ``improvements'' to the flat sky approximation except under very
special circumstances.

We can write the FT of the correlation function, depending on ${\bf u}$ and
${\bf w}$, as
\begin{equation}
  \int d{\bf x}_1\,d{\bf x}_2 \ C({\bf x}_1\cdot{\bf x}_2)
	\exp\left[ 2\pi i {\bf u}\cdot({\bf x}_1-{\bf x}_2)\right]
	\exp\left[ 2\pi i ({\bf w}-{\bf u})\cdot{\bf x}_1\right]
  \quad ,
\end{equation}
where $C(\cos\theta)$ is the (dimensionless) correlation function for the
CMB temperature fluctuations, defined below.  If we expand
\begin{equation}
e^{2\pi i {\bf u}\cdot{\bf x}} = J_0(2\pi u) + 2\sum_{m=1}^\infty
  i^m J_m(2\pi u) \cos\left(m \arccos(\hat{u}\cdot\hat{x})\right)
\end{equation}
and use the symmetry of the problem to do the angular integrals we find
(e.g.~Subrahmanyan et al.~\cite{SESS}) that the diagonal part
\begin{equation}
S(u)\propto \int_0^2 \omega d\omega\ C(\omega) J_0(2\pi u \omega)
\end{equation}
where $\omega=\left|{\bf x}_1-{\bf x}_2\right|=2\sin(\theta/2)$
and $d(\cos\theta)=\omega d\omega$.
In the flat space limit we extend the upper limit of the $\omega$
integration to $\infty$.
Expanding the correlation (or 2-point) function for the CMB temperature
fluctuations as a Legendre series
\begin{equation}
C({\bf x}_1\cdot{\bf x}_2) \equiv \left\langle
  {\Delta T\over T}({\bf x}_1){\Delta T\over T}({\bf x}_2)\right\rangle
= {\displaystyle {1\over 4\pi}} \sum_{\ell=2}^\infty (2\ell+1) C_\ell
	P_\ell({\bf x}_1\cdot{\bf x}_2)
\label{eqn:cordef}
\end{equation}
and using Gradshteyn \& Ryzhik (\cite{GraRyz}; Eq.~7.251(3)) we obtain
\begin{equation}
S(u) = {1\over 2\pi u} \sum_{\ell} (2\ell+1)C_\ell J_{2\ell+1}(4\pi u) .
\label{eqn:suflat}
\end{equation}
In evaluating this sum, care must be taken for regions of the spectrum
where $C_\ell$ is nearly scale invariant due to significant cancellations.

For large $\ell$, $J_{2\ell+1}$ is a sharply peaked function.  Thus the 2D
power spectrum $u^2S(u)\sim\left.\ell(\ell+1)C_\ell\right|_{\ell=2\pi u}$.
Direct numerical evaluation of Eq.~(\ref{eqn:suflat}) or requiring the RMS
fluctuation at zero lag to be the same in $u$ space as $\ell$ space
(Gradshteyn \& Ryzhik \cite{GraRyz}, Eq.~6.511(1)) allows us to write
\begin{equation}
u^2 S(u)\simeq \left. {\ell(\ell+1)\over (2\pi)^2}\ 
  C_\ell \right|_{\ell=2\pi u} \qquad {\rm for}\ u\ga10 \quad.
\label{eqn:suapprox}
\end{equation}
This approximation works at the few percent level for standard CDM (sCDM)
when $u\ga10$ or $\ell\ga60$.

\begin{figure}
\begin{center}
\leavevmode
\epsfysize=6cm \epsfbox{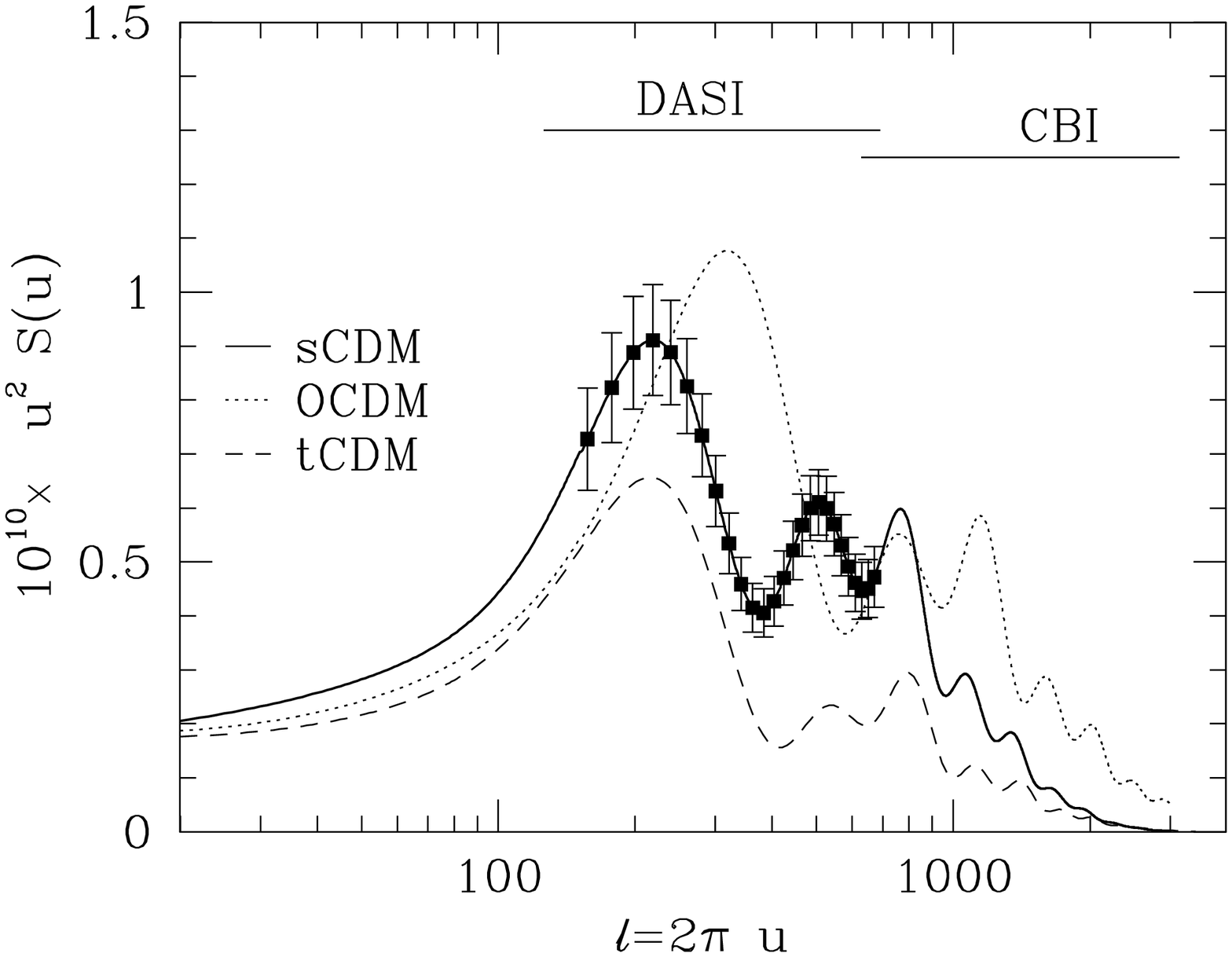}
\epsfysize=6cm \epsfbox{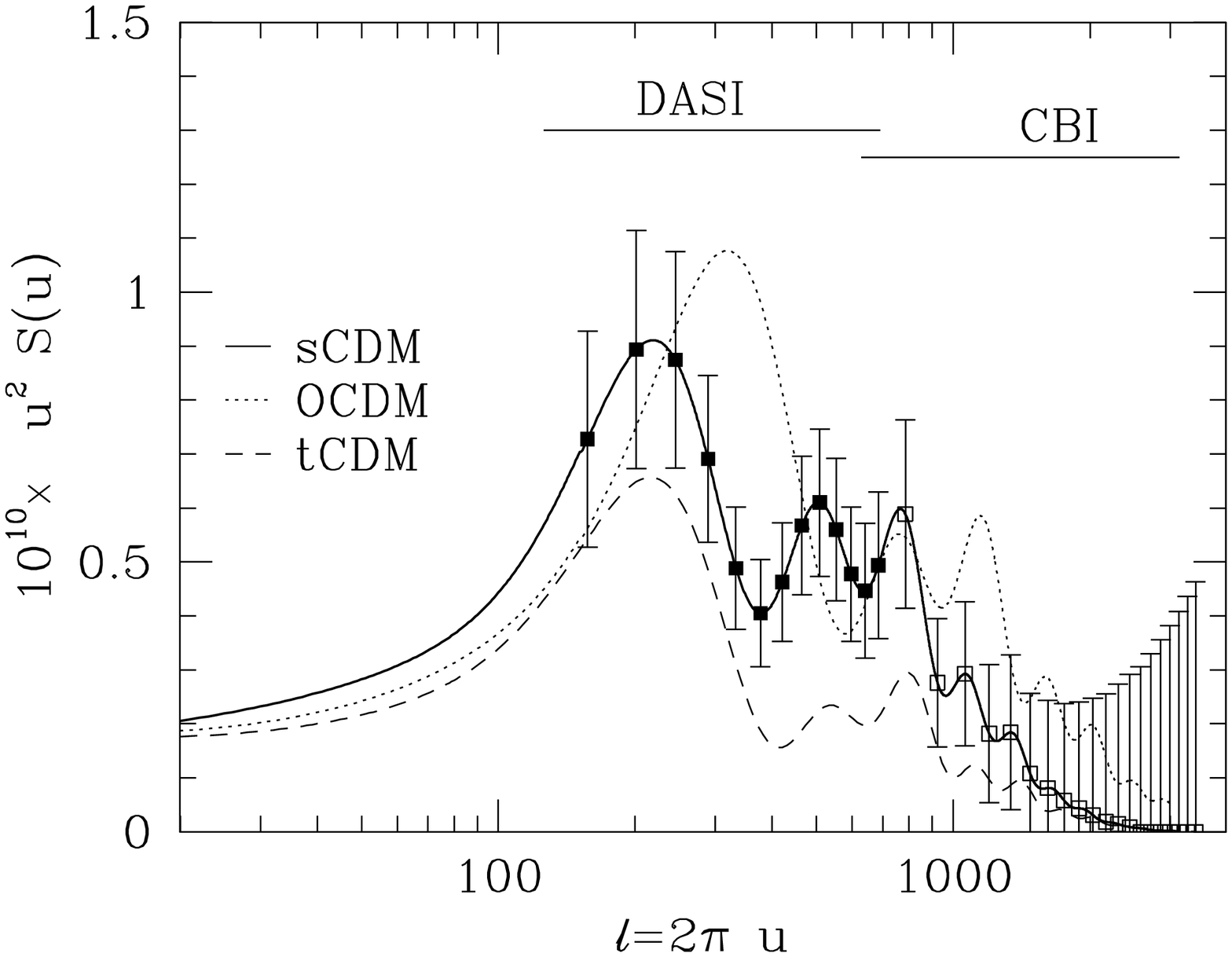}
\end{center}
\caption{The 2D power spectrum per logarithmic interval in $u$, for
a selection of COBE normalized models chosen to provide good fits to
the large-scale structure data (see text).
The solid lines across the top of the plot illustrate the range of scales
to which DASI and CBI will be sensitive.  The left panel shows the expected
uncertainties per bin for the sCDM model for 1 month of observing of 25
widely separated points on the sky.
For display purposes we have placed the points on the sCDM curve.
Each of the 26 baselines is shown and the signal-to-noise is about 1 in
the highest-$u$ bin.
Though each baseline has independent receiver noise, the window functions for
neighboring points overlap considerably so the cosmological signal is very
correlated between points.
The right panel shows the result with 6 months of integration using a
mosaicing strategy covering $\sim400$ square degrees of sky (again with
$S/N\sim1$ in the final bin).  In this case each of these points is completely
independent.  We have also shown the expected error bars for CBI, assuming
a similar mosaicing strategy.}
\label{fig:models}
\end{figure}

We show $S(u)$ vs.~$u$ for a selection of CDM models normalized to the
{\sl COBE} 4-year data in Fig.~\ref{fig:models}.
The range of angular scales that will be probed by DASI and CBI are shown as
the solid lines across the top of the figure.
While these models were chosen primarily to fit the large-scale structure
data, all of the models shown provide reasonable fits to the current CMB data.
The parameters for the models are given in Table~\ref{tab:parms}.

\subsection{The Visibility Correlation Matrix}

In theories which predict gaussian temperature fluctuations the fundamental
theoretical construct is the correlation matrix of the measured data.  Since
the data in our case are the visibilities measured at a set of points
${\bf u}_i$, in what follows we will need to know the correlation matrices for
the signal and noise in the various visibilities.
The measured fluxes, $V({\bf u})$, are
\begin{eqnarray}
V({\bf u}) = {\partial B_\nu\over\partial T} T_{\rm CMB} \int d{\bf x}
  \ {\Delta T({\bf x})\over T_{\rm CMB}}\ A({\bf x})
  e^{2\pi i{\bf u}\cdot{\bf x}}
\end{eqnarray}
where ${\partial B_\nu/\partial T}$ converts from temperature to intensity,
$T_{\rm CMB}$ is the CMB temperature and $A({\bf x})$ is the primary beam.
The conversion factor from ``temperature'' to ``intensity'' is
\begin{equation}
{\partial B_\nu\over\partial T} =
  {2k_B\over c^2} \left( {k_BT\over h} \right)^2 {x^4 e^x\over (e^x-1)^2}
  \simeq \left( {99.27\,{\rm Jy}\ {\rm sr}^{-1}\over \mu\,{\rm K}} \right)
  {x^4 e^x\over (e^x-1)^2}
\end{equation}
where $B_\nu$ is the Planck function, $k_B$ is Boltzman's constant,
$x\equiv h\nu/k_BT_{\rm CMB}\simeq\nu/56.84\;$GHz is the
``dimensionless frequency'' and $1\,$Jy=$10^{-26}\,$W/m${}^2$/Hz.
In the Rayleigh-Jeans limit
${\partial B_\nu/\partial T}\simeq 2k_B(\nu/c)^2$ where $\nu$ is the
observing frequency.
This is a good approximation for the frequencies of planned interferometers,
the correction to the Rayleigh-Jeans assumption is 2\% at $30\;$GHz, and we
shall make it henceforth.  We shall also set $c\equiv 1$ so $\nu=\lambda^{-1}$.

The Fourier Transform of the primary beam\footnote{Throughout we will use
a tilde to represent the Fourier Transform of a quantity.} is the
auto-correlation of the Fourier Transform of the point response, $g$, of the
receiver to an electric field,
$\widetilde{A}(u)=\widetilde{g}\star\widetilde{g}(u)$ and
\begin{equation}
A({\bf x})=\int d{\bf u}\ \widetilde{A}({\bf u})e^{-2\pi i{\bf u}\cdot{\bf x}}
\quad ,
\end{equation}
so using the fact that the power spectrum is diagonal in ${\bf u}$ we have
\begin{equation}
C_{ij}^{V} \equiv \left\langle V^*({\bf u}_i)V({\bf u}_j)\right\rangle =
    \left({2k_BT_{\rm CMB}\nu^2}\right)^2
    \int d^2w\ \widetilde{A}^*({\bf u}_i-{\bf w})
    \widetilde{A}({\bf u}_j-{\bf w})S(\left|{\bf w}\right|) \quad .
\label{eqn:cvdef}
\end{equation}
Notice that the visibilities are uncorrelated if $|{\bf u}_i-{\bf u}_j|$
is larger than (twice) the width of $\widetilde{A}$, which defines the bin
size $\Delta u$.

We are now in a position to define the window function $W_{ij}(u)$ which,
when convolved with the power spectrum, defines the visibility correlation
matrix $C_{ij}^{V}$.  From the above
\begin{equation}
C_{ij}^{V} = \left({2k_BT_{\rm CMB}\nu^2}\right)^2
 \int_0^\infty w dw\ S(w) W_{ij}(w)
\label{eqn:suwij}
\end{equation}
with
\begin{equation}
W_{ij}(|{\bf w}|) \equiv
    \int_0^{2\pi} d\theta_{w}\ \widetilde{A}^*({\bf u}_i-{\bf w})
    \widetilde{A}({\bf u}_j-{\bf w}) \quad .
\label{eqn:wijdef}
\end{equation}
Note that since $\widetilde{A}$ has compact support the maximum of
$W_{ii}$ scales as $u_i^{-1}$ for $u_i\gg \Delta u$.
Since the noise per visibility is independent of $u_i$ the signal-to-noise
drops as $u^{-2}$ for a scale-invariant spectrum, $S(u)\propto u^{-2}$.
Also note that both $W_{ij}$ and $S(u)$ are positive semi-definite, so
the visibilities are never anti-correlated, unlike single-dish (chopping)
experiments.

In general the distribution of the electric field in the horn aperture is
close to a pillbox times a Gaussian.  Due to the finite aperture
$\widetilde{A}$ has compact support.
Usually the response is independent of $\hat{u}$:
we will call the $|{\bf u}|$ for which $\widetilde{A}$ vanishes $\Delta u$.
In order to obtain a simple estimate of our window function it is a
reasonable first approximation to take $\widetilde{A}$ equal to the
auto-correlation of a pillbox of radius $D/2$ where $D$ is the diameter of
the dish.  Specifically
\begin{equation}
\widetilde{A}({\bf u}) =
  {2\widetilde{A}_*\over\pi} \left[ \arccos{u\over D} -
  {u\sqrt{D^2-u^2}\over D^2} \right]
\label{eqn:tildeA}
\end{equation}
if $u\le D$ and zero otherwise.  Thus in this simple example $\Delta u=D$.
If we require $A(0)=1$ then this must integrate to unit area, so
$\widetilde{A}_*^{-1}=\pi(D/2)^2$, or the area of the dish.

\begin{figure}
\begin{center}
\leavevmode
\epsfysize=6cm \epsfbox{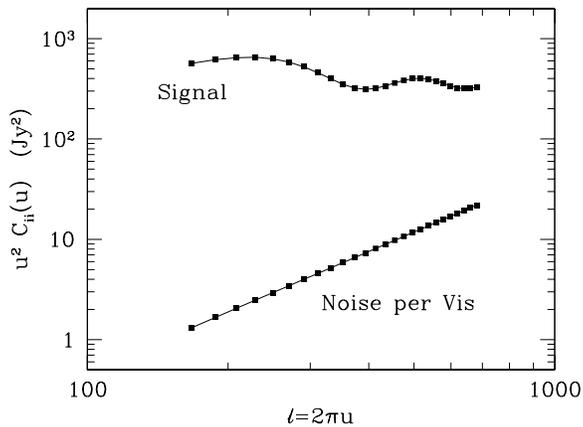}
\end{center}
\caption{The 26 diagonal elements of the correlation matrix $C_{ii}^V$ and
$C_{ii}^N$ for {\sl COBE\/} normalized sCDM and a pillbox aperture function.
The noise is assuming 1 day of operation of DASI and is per visibility.
We have shown results for 1 pointing, without the instrument rotation
necessary to fill the $u-v$ plane.
The signal in each of these visibilities will be highly correlated due to
strongly overlapping window functions (see text).}
\label{fig:corrfunction}
\end{figure}

In the case where all correlated signal is celestial, the correlation function
of the noise is diagonal with
\begin{equation}
C^N_{ij}=\left( {2k_BT_{\rm sys}\over\ {\eta_A A_D} } \right)^2
 {1\over \Delta_\nu t_a n_b}\ \delta_{ij} \quad .
\label{eqn:cndef}
\end{equation}
Here $T_{\rm sys}$ is the system noise temperature, $\lambda$ the wavelength,
$\eta_A$ is the aperture efficiency, $A_D$ is the physical area of a dish (not
to be confused with $A({\bf x})$), $n_b$ is the number of
baselines\footnote{The number of baselines formed by $n_r$ receivers is
$n_b=n_r(n_r-1)/2$.} corresponding to a given separation of antennae,
$\Delta_\nu$ is the bandwidth and $t_a$ is the observing time.
Typical values for DASI are $T_{\rm sys}=20$K, $\eta_A\sim0.8$, dishes
of diameter $20\;$cm, $n_b=3$ and $\Delta_\nu=10\;$GHz
(in $10\times1\;$GHz channels).
For CBI the dishes are 5 times larger with the other numbers about the
same.
We show in Fig.~\ref{fig:corrfunction} the diagonal entries of $C^V$ and
$C^N$ for one pointing and 1 day of observing with DASI.

Using Eqs.~(\ref{eqn:cvdef}, \ref{eqn:cndef}) we can provide a rough estimate
of the signal-to-noise expected in a given visibility.  For $\widetilde{A}$
given by Eq.~(\ref{eqn:tildeA}) we have
\begin{equation}
  {C^V_{ii}\over C^N_{ii}} \sim
  \left( {T_{\rm CMB}\over T_{\rm sys}} \right)^2
  \Delta_\nu t_a n_b\ \left( {\Delta u\over u} \right)^2
  \ u^2 S(u)
\end{equation}
which given the numbers above yields
$C^V_{ii}/C^N_{ii}\sim 10^3 \left( \Delta u/u\right)^2$ for 1 day of
integration, assuming a COBE normalized, scale-invariant spectrum.

\section{Comparison with Theory} \label{sec:comparetheory}

\subsection{Likelihood function}

In stochastic theories $\Delta T(\hat{x})$ is a (gaussian) random variable
with zero mean and dispersion given by the $C_\ell$.  Thus the visibilities
measured by the interferometer will be gaussian random variables with zero
mean and dispersion $C^V+C^N$.
For a given set of measured visibilities one can test any theory, or
set of $\{C_\ell\}$, by constructing the likelihood function (for complex
variables $V$)
\begin{equation}
{\cal L}\left( \{C_\ell\}\right) = {1\over\pi^n \det{C}} \ \exp
  \left[ -V^{*}({\bf u}_i) C_{ij}^{-1}V({\bf u}_j)\right]
\label{eqn:likelihood}
\end{equation}
where $C_{ij}=C^V_{ij}+C^N_{ij}$ is the correlation matrix of visibilities
at ${\bf u}_i$ and ${\bf u}_j$ (Hobson, Lasenby \& Jones \cite{HobLasJon}).
Note that the visibilities are complex and thus the likelihood function is
slightly different than for the case of real gaussian random variables,
e.g.~the ``missing'' factor of ${1\over 2}$ in the exponent.
The restriction to the half-plane however ensures that the number of degrees
of freedom is the same as for the real case
(c.f.~Hobson, Lasenby \& Jones \cite{HobLasJon}).

\begin{table}[htb]
\begin{center}
\begin{tabular}{lcccc}
Name & $\Omega_{\rm mat}$ & $h$ & $\Omega_{\rm B}h^2$ & $n$ \\ \hline
sCDM &   1  & 0.5  &  0.0125 & 1 \\
OCDM & 0.5  & 0.6  &  0.0200 & 1 \\
tCDM &   1  & 0.5  &  0.0250 & 0.8
\end{tabular}
\end{center}
\caption{The cosmological parameters for the theories discussed in the text.
All models have $\Omega_{\Lambda}=0$ and have been normalized to the
{\sl COBE} DMR 4-year data.}
\label{tab:parms}
\end{table}

Given a set of data $\{V_i\}$ one can proceed to test theories using the
likelihood function.  Confidence intervals for parameters and relative
likelihood of theories are calculated in the usual way.

\subsection{Power Spectrum Estimation} \label{sec:quadratic}

There are several ways one could consider estimating the angular power
spectrum from a set of visibilities.  Conceptually the simplest is to
average $|V(\vec{u})|^2$ in shells of constant $|\vec{u}|$. This gives a
(noised biased) estimate $C_\ell$ at $\ell=2\pi u$, convolved with the
window function $W_{ij}$.

A more sophisticated method is to define as a ``theory'' a set of
bandpowers, i.e.~define the power spectrum $\ell(\ell+1)C_\ell$ as a
piecewise constant, in $N_{\rm band}$ bands $B$.  One then maximizes
the likelihood function for this ``theory'', the result is the best
fitting power spectrum, binned into groups at similar $\ell$.
Such an approach has been used on the {\sl COBE\/} 4-year data by
Gorski~(\cite{Goretal}) and Bunn \& White~(\cite{BunWhi}), and discussed
extensively by Bond, Jaffe \& Knox~(\cite{BonJafKno}).
It is the method we shall advocate here.

As pointed out by Bond, Jaffe \& Knox~(\cite{BonJafKno}), it is particularly
simple to find the maximum of the likelihood function for such a ``theory''
using a quadratic estimator (see also Tegmark~\cite{Max}) which takes as an
input a trial theory.  In this case the theory is a set of ``bandpowers''
chosen to cover the $\ell$ range of interest and be approximately independent.
Iteration of the quadratic estimator is equivalent to Newton's method for
finding the root of $d{\cal L}/dp$ where $p$ is a parameter on which
${\cal L}$ depends.

Specifically from an estimate of the bandpowers,
$u^2S(u)=\hat{p}_\alpha$ for $u\in B_\alpha$
(where $\alpha=1,\cdots,N_{\rm band}$) an improved estimate is
$p_\alpha=\hat{p}_\alpha+\delta p_\alpha$ with
\begin{equation}
\delta p_\alpha = \left( {\rm tr}\left[
 \widehat{C}^{-1}\widehat{C}_{,\alpha}\
 \widehat{C}^{-1}\widehat{C}_{,\beta}\right]\right)^{-1}
 {\rm tr}\left[ \left( C-\widehat{C} \right)
 \left( \widehat{C}^{-1}\widehat{C}_{,\beta}\widehat{C}^{-1}\right) \right]
\end{equation}
(Bond, Jaffe \& Knox~\cite{BonJafKno}).
Here $\widehat{C}$ indicates the (theoretical) correlation matrix evaluated at
the initial estimate $\hat{p}_\alpha$ with $C_{,\alpha}\equiv dC/dp_\alpha$
and $C\equiv V^{*}({\bf u}_i)V({\bf u}_j)$ indicates the matrix formed by the
data.  Iteration of this procedure (e.g.~from an initially flat spectrum
$p_\alpha=$constant) converges to the maximum likelihood estimate of the
power spectrum.

Since the parameters $p_\alpha$ are chosen to be the (constant) values of
$u^2S(u)$ across the band $B_\alpha$ and the noise is assumed to be
independent of the level of cosmological signal
\begin{equation}
\left[ \widehat{C}_{,\alpha} \right]_{ij}
  =  \left({2k_BT_{\rm CMB}\nu^2}\right)^2
     \int_{B_\alpha} {dw\over w}\ W_{ij}(w)
\quad .
\end{equation}

We may gain some intuition for this expression by considering the simple
case of 1 band and uncorrelated visibilities.  Assume additionally that
$W_{ij}({\bf u})$ is independent of $\hat{u}$.  Then if we write
$\widehat{C}=\sigma_i^2 w_i \delta_{ij}$
and $\widehat{C}_{,\alpha}=w_i\delta_{ij}$ we have
\begin{equation}
\delta p_\alpha = \left( \sum_i \sigma_i^{-4} \right)^{-1}
  \sum_j {|V_j|^2 w_j^{-1} - \sigma_j^2\over \sigma_j^4 }
\quad .
\end{equation}
This becomes even simpler if all of the visibilities are at fixed
$|{\bf u}_i|$.  Then $\sigma_i^{V}$ is independent of $i$ and, if we assume
the noise is also, $\delta p$ becomes zero when
\begin{equation}
  \left( \sigma^{V} \right)^2 =
  {1\over N} \sum_{j=1}^{N} |V_j|^2 w_j^{-1} - \left( \sigma^{N} \right)^2
\end{equation}
which is reminiscent of our simplistic estimate described at the beginning
of this section.

\subsection{CDM Parameter Estimation}

It has become common
(Scott \& White~\cite{Echoes}, Jungman et al.~\cite{JKKS},
Bond, Efstathiou \& Tegmark~\cite{BET},
Zaldarriaga et al.~\cite{ZalSpeSel}, Stompor \& Efstathiou~\cite{StoEfs})
to ask how well we could measure theory parameters ``on average'' given a set
of measurements $\{V_i\}$ which are ``typical''.
Imagine that our theory is defined by a set of parameters $\{p_\alpha\}$ and
that the sky corresponds to this theory with values of parameters
$\hat{p}_\alpha$.  In this case
\begin{equation}
\left\langle -\ln {\cal L} \right\rangle =
  {\rm tr}\left[ \widehat{C}C^{-1}+\ln C \right]
\end{equation}
where $\widehat{C}_{ij}=\langle V^{*}_iV_j\rangle$ denotes $C_{ij}(\hat{p})$.
The precision to which we can measure $\{p_\alpha\}$ assuming
an input theory $\widehat{C}$ is given by the second derivative matrix of
$\left\langle -\ln {\cal L}\right\rangle$
\begin{equation}
\left.
{\partial^2\langle -\ln{\cal L}\rangle\over \partial p_\alpha \partial p_\beta}
\right|_{\hat{p}}
= {\rm tr}\left[
 \widehat{C}^{-1}\widehat{C}_{,\alpha}\
 \widehat{C}^{-1}\widehat{C}_{,\beta}\right]
\end{equation}
with $C_{,\alpha}\equiv dC/dp_\alpha$. Thus
\begin{equation}
\left\langle
(p_\alpha-\langle p_\alpha\rangle) (p_\beta -\langle p_\beta\rangle)
\right\rangle_{\cal L} = \left\{ {\rm tr}\left[
 \widehat{C}^{-1}\widehat{C}_{,\alpha}
\  \widehat{C}^{-1}\widehat{C}_{,\beta}\right]
\right\}^{-1}
\label{eqn:twopoint}
\end{equation}
where $\langle\cdots\rangle_{\cal L}$ denotes an average with respect to
the likelihood function ${\cal L}$.

As an example of how well planned interferometers will constrain cosmological
models we can ask how well DASI would be able to determine the cosmological
parameters $p_\alpha$=
\{$C_{10}$, $\Omega_{\rm mat}$, $h$, $\Omega_{\rm B}h^2$, $n$\}
for the adiabatic CDM models listed in Table~\ref{tab:parms}.
We have not included the ionization history in this list since the amplitude
and the ionization history are degenerate on these scales for late
reionization.  We will also assume that $\Omega_\Lambda=0$.
We choose the value of $C_\ell$ at $\ell=10$ for our normalization parameter
as this is approximately the ``pivot point'' of the {\sl COBE\/} data.

In order to avoid a detailed modeling of the DASI observing strategy before
the instrument is operational, we shall simply assume that DASI will
fully sample the $u-v$ half-plane between $u=25$ and 110.
The number of different $\hat{u}$ needed to cover the $u-v$ ring at $u$ if
the aperture has a radial width $2\Delta u$ is $N(u)=2u/\Delta u$, where we
have made use of the fact that only the {\it half\/}-plane is covered.
(This is actually conservative -- points do not become correlated as soon as
the $\widetilde{A}$'s touch.  We would want to sample a factor of $\sim2$
more densely than this and keep track of correlations, but we shall postpone
an optimization to a later paper).
The DASI has continuous sensitivity in $|{\bf u}|$.
Thus we assume that we can fully cover the $u-v$ plane with $N(u_{\rm max})$
pointings (orientations) of the instrument.
We spend equal time on each orientation.

To simplify this computation we will count only the {\it independent\/}
visibility measurements arising from such a strategy, i.e.~ignore the extra
information that will be available from the correlations.
Specifically we choose ${\bf u}_i$ such that there are $N_{\rm bin}$
independent bins in the radial ($|{\bf u}|$) direction and $N(|{\bf u}_i|)$
bins in the angular ($\hat{u}$) direction.  Eq.~(\ref{eqn:twopoint}) in the
case that $C_{ij}$ is diagonal reduces simply to
\begin{equation}
\left\langle
(p_\alpha-\langle p_\alpha\rangle) (p_\beta -\langle p_\beta\rangle)
\right\rangle_{\cal L}^{-1} = \sum_{i=1}^{N_{\rm bin}} N(|{\bf u}_i|)
{\widehat{C}^V_{ii,\alpha}\widehat{C}^V_{ii,\beta} \over
\left( \widehat{C}_{ii}^{V} + \widehat{C}_{ii}^{N} \right)^2 }
\end{equation}
where we have assumed the noise is independent of $p_\alpha$.
The width of the aperture $\Delta u$ is chosen so that
$u_{\rm max}-u_{\rm min}=2N_{\rm bin}\Delta u$.  To obtain this resolution
in $u$ requires increasing the sky coverage either by shrinking the dish size
or by mosaicing (see \S\ref{sec:mosaic}).  We shall assume that if mosaicing
is used then the spherical symmetry of $\widetilde{A}$ is preserved.  As
discussed in \S\ref{sec:mosaic}, this will only be an approximation.
A full analysis of the strategy including mosaicing, which takes into account
the deviations from the monochromatic and ``flat sky'' approximations and
includes the correlations between visibilities remains to be done.
The number of bins is used here as a measure of the sky coverage achieved
by the experiment.  For reference $N_{\rm bin}=20$ corresponds to a
$10^3$ square degree circle of sky.

We shown in Table~\ref{tab:errors} the relative uncertainties on $p_\alpha$,
with a ``prior'' 20\% uncertainty in $C_{10}$.  (We do not include the increase
in the errors from foreground subtraction here.)
The number of bins is fixed at $N_{\rm bin}=20$, or $10^3$ square degrees of
sky coverage, for simplicity.
For 3 months of observing an sCDM sky the DASI would be system noise limited
for $u\ga50$.
Note that there is a strong dependence of the estimated errors on the input
theory (and the parameter set chosen).  We have chosen the 3 theories here
to explore this dependence.
For the open model (which has $\Omega_0=0.5$) the peaks in the power
spectrum are at higher $u$ than the critical density models
(see Fig.~\ref{fig:models}).
This means that there is less information (from the higher peaks) available
to break the degeneracy in parameter variations, leading to larger
uncertainties when the other parameters are integrated out.
To increase the precision with which cosmological parameters could be
determined in an open model, one would need to extend the coverage to higher
$|{\bf u}|$ using CBI.  If this is done the angular scale of the features in
the open model is better matched to DASI+CBI than in the flat models, and the
parameters can thus be better determined.
For the tilted model the signal-to-noise is lower at large $u$, which accounts
for the slightly larger uncertainties on e.g.~the spectral slope $n$.
However the higher baryon fraction leads to greater sensitivity to
$\Omega_{\rm B}$, and the variations with $\Omega_{\rm mat}$ are less
correlated with $\Omega_{\rm B}$.

\begin{table}[htb]
\begin{center}
\begin{tabular}{l|cccc|cccc|cccc} \hline
 & \multicolumn{4}{c|}{3 months} & \multicolumn{4}{c|}{6 months}
 & \multicolumn{4}{c}{12 months} \\ \hline
Name & $\Omega_{\rm mat}$ & $h$ & $\Omega_{\rm B}$ & $n$
     & $\Omega_{\rm mat}$ & $h$ & $\Omega_{\rm B}$ & $n$
     & $\Omega_{\rm mat}$ & $h$ & $\Omega_{\rm B}$ & $n$
\\ \hline
sCDM & 21 & 40 & 50 &  9 & 15 & 26 & 34 &  7 & 12 & 19 & 27 &  5 \\
OCDM & -- & -- & -- & 62 & 73 & -- & -- & 45 & 53 & -- & 91 & 33 \\
tCDM & 30 & 71 & 40 & 16 & 17 & 39 & 25 & 10 & 11 & 23 & 17 &  7
\end{tabular}
\end{center}
\caption{The {\it percentage\/} relative uncertainties for the cosmological
parameters assuming 3, 6 or 12 months of integration of the DASI over $10^3$
square degrees ($N_{\rm bin}=20$), for the theories discussed in the text.
We have included a ``prior'' corresponding to 20\% uncertainty in $C_{10}$.
For the open models several parameters are very correlated because DASI
alone does not probe to high enough $u$.
For sCDM with $N_{\rm bin}=20$ and $T_{\rm sys}=20$K we are noise limited for
$u\ga50$ with 3 months observing time.
Once DASI and CBI are combined the uncertainties on all parameters for all
models are $<20\%$ for 1 year of data.}
\label{tab:errors}
\end{table}

For all of these theories the uncertainties are larger due to the large
correlations between parameters:
if the bin size or noise is too large then one cannot distinguish between
variations in different parameters, inflating the marginalized errors.
To decrease the correlations one must increase the resolution in $u$, which
means obtaining more sky coverage, or increase the total range of $u$ covered,
which means combining DASI and CBI.
The combination of DASI and CBI is particularly powerful, with the marginalized
errors for each model being below 20\% for all parameters within 1 year
of observing.

\section{Mosaicing} \label{sec:mosaic}

\subsection{Increasing Resolution}

The resolution which we have in $u$-space is limited by the amount of sky that
we have surveyed, which for a single pointing is equal to the size of the
primary beam.
By combining several contiguous pointings of the telescope we can increase
the amount of surveyed sky and therefore increase the resolution in $u$-space.
This is known as mosaicing.
The idea here is to measure the visibility as a function of position
${\bf y}$ from the map or phase center
\begin{equation}
V_{{\bf y}}({\bf u}) \propto \int d^2x\ A({\bf x}-{\bf y}) T({\bf x})
  e^{2\pi i {\bf u}\cdot{\bf x}}
\end{equation}
We then sample $V_{{\bf y}}$ at a series of points by repointing the entire
telescope.
Let us denote this by a sampling function $\Pi({\bf y})$ which will be a sum
of delta functions.
We compute the Fourier Transform of $\Pi({\bf y})V_{{\bf y}}$ which is
simply the convolution $\{\widetilde{\Pi}\star\widetilde{V}\}({\bf v})$
where
\begin{equation}
\widetilde{V}({\bf u},{\bf v}) =
  \widetilde{A}({\bf v}) \widetilde{T}({\bf u}+{\bf v}) \quad .
\label{eqn:vuv}
\end{equation}
Here ${\bf u}$ is the original vector in the $u-v$ plane, and ${\bf v}$ is
the variable conjugate to ${\bf y}$.

\begin{figure}
\begin{center}
\leavevmode
\epsfysize=6cm \epsfbox{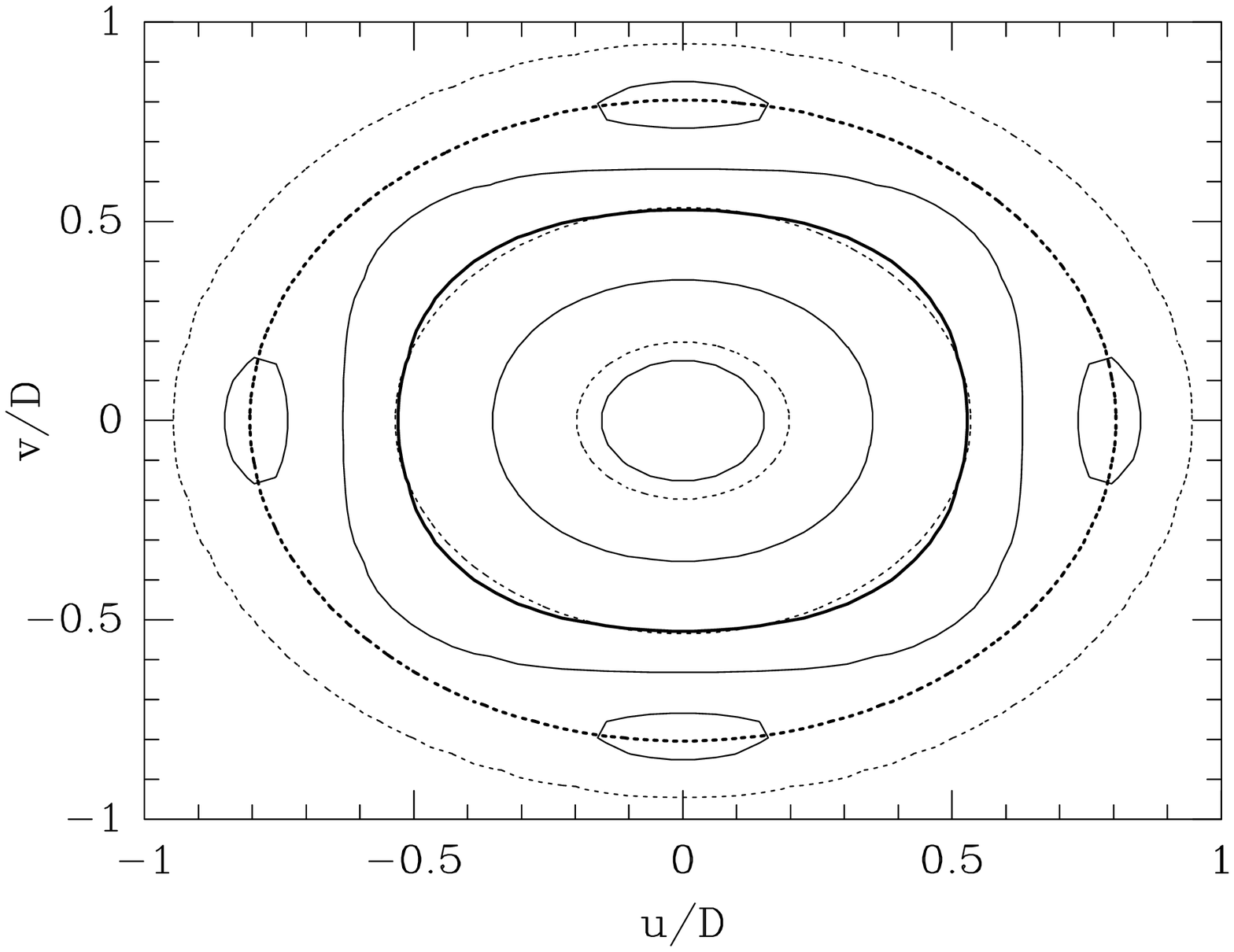}
\epsfysize=6cm \epsfbox{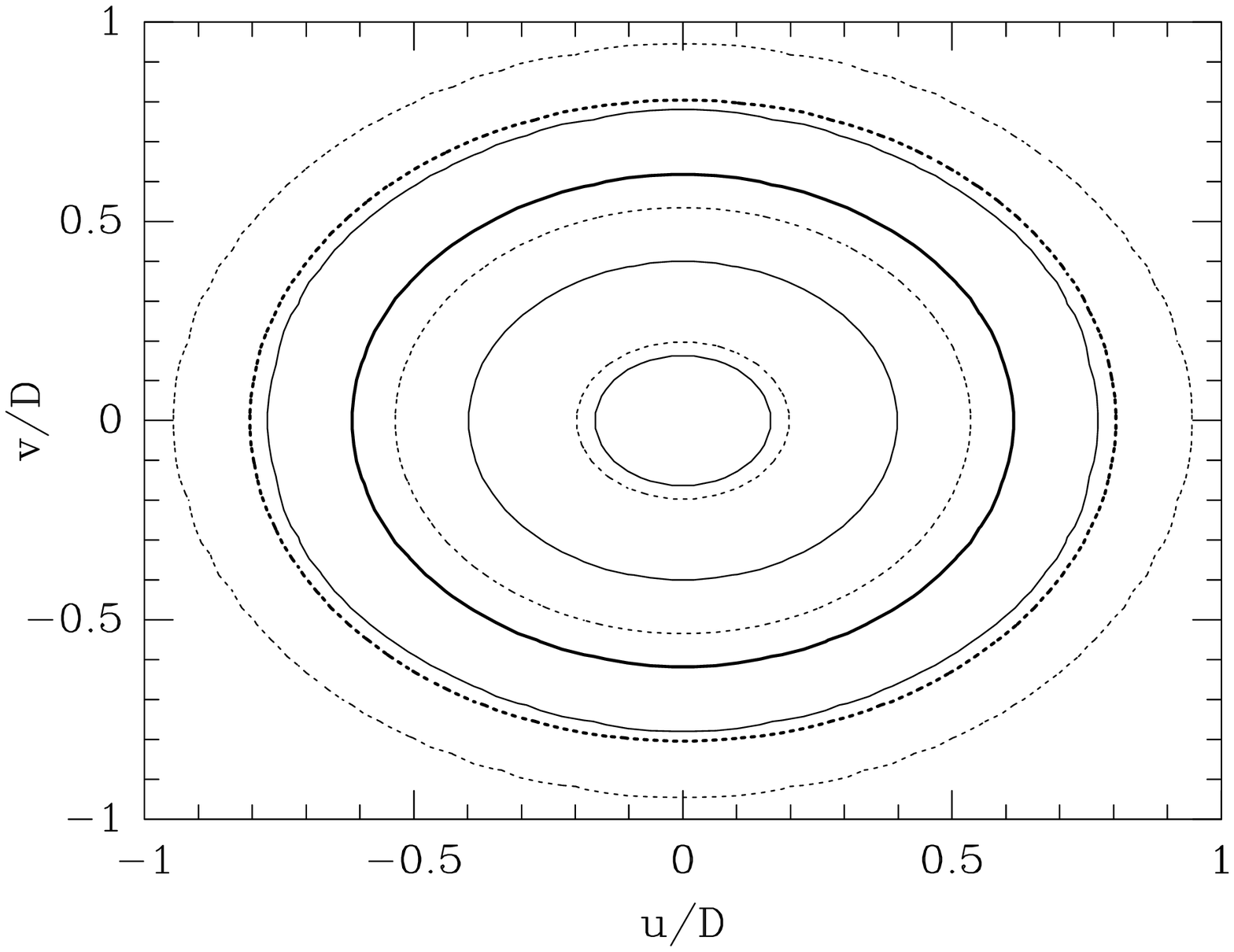}
\end{center}
\caption{The aperture function $|A(\vec{u})|^2$ for (a) $3\times3$ square
mosaic with spacing $D/2$ (b) 7 points laid out in 3 rows (at the center and
vertices of a regular hexagon) with all neighbors separated by $D/2$.
Contours mark 1, 2, 3 and $4\sigma$ from the peak with $3\sigma$ shown thick
($1\sigma=e^{-1/2}$ of peak power).
The dashed lines indicate the unmosaiced aperture of the primary beam, the
solid lines the result after mosaicing.}
\label{fig:mosaic}
\end{figure}

If our sampling function is a sum of delta functions then $\widetilde{\Pi}$
is a sum of plane waves.  Thus simply summing our $V_{{\bf y}}$ at points
${\bf y}_j$ with weight $c_j$ changes our aperture in $u$-space from
$\widetilde{A}$ to
\begin{equation}
 \sum_j c_j e^{-2\pi i{\bf v}\cdot{\bf y}_j}\ \widetilde{A}({\bf v})
\end{equation}
which can be made much narrower.
This is completely analogous to the usual case of Fraunhofer diffraction
through many holes, treated in most textbooks on optics.
The simplest example is when $c_j=1$ and the ${\bf y}_j$ lie on a regular
$N\times N$ grid with spacing ${\bf \delta}$ then
\begin{equation}
 \sum_j e^{-2\pi i{\bf v}\cdot{\bf y}_j} =
 \exp\left[-2\pi i {\bf v}\cdot{\bf \delta}\ {(N-1)\over 2}\right]\
 {\sin\left[N 2\pi v_x\delta_x\right]\over\sin\left[2\pi v_x\delta_x\right]}\
 {\sin\left[N 2\pi v_y\delta_y\right]\over\sin\left[2\pi v_y\delta_y\right]}
 \quad .
\end{equation}
The second diffraction spike occurs at $v_{x,y}=\delta_{x,y}^{-1}$.  If
we choose $\delta_{x,y}$ sufficiently small (Nyquist sampling) then this will
be outside the range where $\widetilde{A}$ vanishes\footnote{If we sample at
precisely the spatial Nyquist rate, gain variations from stare to stare will
also alias power outside of the spatial frequency window.}.
Fig.~\ref{fig:mosaic} shows the gain for a $3\times3$ mosaic.

In the absence of a preferred direction in the theory, the power spectrum of
the sky is symmetric in $\hat{u}$, so choosing a mosaicing strategy based on
a cartesian grid is not optimal.  We would like to have the mosaicing maximally
rotationally symmetric within the observing constraints.
In Fig.~\ref{fig:mosaic} we compare the $3\times3$ square to 7 points laid
out at the center and vertices of a regular hexagon, with all points
equidistant from their neighbors.

Note that mosaicing does not increase the {\it range\/} of $u$ to which we
are sensitive (see Eq.~\ref{eqn:vuv}), it simply enhances our
{\it resolution\/} by allowing us to follow more periods of a given wave.
Thus we retain our ability to reject long-period noise or foregrounds.

If the goal were simply a measurement of the power spectrum, it is just as
efficient to use smaller telescopes (with intrinsically better $u$ resolution)
and integrate for longer as it is to use large dishes and mosaic to increase
the resolution. The advantage of the later method is that it allows better
imaging of each piece of sky for checks of systematics, non-gaussian features
and foregrounds and it is easier to avoid ``dropouts'' in the $u$ coverage
for the power spectrum.

There are two routes to analyzing mosaiced data.  The first is to treat
the visibilities $V_{{\bf y}_j}({\bf u}_i)$ as separate data, highly
correlated in a calculable way, with apparently low resolution but much
information in the correlations between visibilities.
The theory correlation matrix of mosaiced data can be written in the
form of Eq.~(\ref{eqn:suwij}) with a modified window function which allows
different pointing centers for each visibility
\begin{equation}
W_{ij}(|{\bf w}|) \equiv
    \int_0^{2\pi} d\theta_{w}\ \widetilde{A}^*({\bf u}_i-{\bf w})
    \widetilde{A}({\bf u}_j-{\bf w})
    e^{ 2 \pi i \left[ ({\bf u}_j-{\bf w})\cdot {\bf y_j}
    -({\bf u}_i-{\bf w})\cdot{\bf y_i} \right] } \quad .
\end{equation}
We discuss this further in \S\ref{sec:osf} below.

The other route is to statistically weight the $V({\bf u}_i)$ from the
different ${\bf y}_j$ to form a synthesized data set with fewer visibilities
and correlations and intrinsically higher resolution.
This method is perhaps simpler to understand but the weighting of the
different ${\bf y}_j$ will probably not be optimal for parameter estimation
or power spectrum estimation.

\subsection{Optimal Subspace Filtering} \label{sec:osf}

A set of data from several (mosaiced) pointings involves many measured
visibilities, very few of which are independent.  Thus likelihood analysis
requires repeated inversion of a large matrix, which can be computationally
quite slow.  One method for increasing the efficiency of the calculation is
to work with a subset of the data which contains most of the signal, but has
fewer elements.  This transformation can be accomplished using an ``optimal
subspace filter''.

Consider first the case of a single (unmosaiced) pointing.
For the {\sl DASI\/} the visibilities will be measured at a specified set of
${\bf u}$ at any one time, and then the instrument will be rotated about an
axis through the center of the aperture plane to obtain a different set of
${\bf u}$ with the same lengths but different orientations.
In this way most of the $u-v$ plane will be covered allowing for imaging.
The number of rotations will be chosen to almost fully sample the $u-v$
plane at the highest resolution (largest baseline or largest $u$).
This means that considerable oversampling of the smaller baselines must
result, since the aperture $\widetilde{A}$ has a fixed width $\Delta u$.
Phrased another way, the correlation matrix $C_{ij}^V$ of our visibilities
has dimension $\sim(u_{\rm max}/\Delta u)^2$.
However only a fraction of the entries are uncorrelated and contain ``new''
information about the theory.  The rest are primarily a measure of the noise
in the experiment.
Thus if we could perform a change of basis to those combinations which
measure primarily signal and those which measure primarily noise we could
work in a much smaller subspace of the data (the ``signal'' subspace) with
little loss of information about the theory.
In calculating the likelihood function we need to invert $C_{ij}$.
Since matrix inversion is an $N^3$ process, reducing the amount of data
can dramatically decrease the processing time.

As a simple example of this technique, imagine that we measure a visibility
twice, with independent noise in each measurement.  The signal in these
two measurements is totally correlated.
The sum of the two measurements is primarily sensitive to the signal
(with noise $1/\sqrt{2}$ of each individual measurement) while the difference
measures primarily noise.  Thus if we ``change basis'' to the sum and
difference of the two measurements we could work with the sum only with little
loss of information.  This would halve the size of the matrix we would need
to invert in the likelihood function and speed processing by a factor $\sim8$.

Optimal subspace filtering (also known as the
{\it signal-to-noise eigenmode analysis\/} or
{\it Karhunen-Loeve transform\/})
is a method designed to estimate the best change of basis and subspace
in which to work.
Recent applications of this method to the CMB and large-scale structure can
be found in
(Bond~\cite{Bond}, Bunn \& Sugiyama~\cite{BunSug},
Bunn, Scott \& White~\cite{BunScoWhi}, White \& Bunn~\cite{WhiBun},
Bunn \& White~\cite{BunWhi}, Vogeley \& Szalay~\cite{VogSza},
Bond \& Jaffe~\cite{BonJaf}, Tegmark, Taylor \& Heavens~\cite{TegTayHea}
and Tegmark, et al.~\cite{TegHamStrVogSza}).
In the above example $C_{ij}=1$ for all $i,j$.  This matrix has eigenvalues
2 and 0 and (orthonormal) eigenvectors $(1,1)/\sqrt{2}$ and $(1,-1)/\sqrt{2}$
respectively.  The eigenvalues measure the signal-to-noise carried by the
combination of data points $\vec{d}\cdot\vec{\psi}_a$, where $\vec{\psi}_a$
are the (orthonormal) eigenvectors.

In general it can be shown that to find the optimal subspace on which to
project one finds the eigenvalues and eigenvectors of the matrix
$\sigma_i^{-1}C_{ij}\sigma_j^{-1}$ where $\sigma_i$ is the noise\footnote{If
the noise is not diagonal then the appropriate matrix is $N^{-1/2}CN^{-1/2}$
where $N^{1/2}$ is any square root of the noise correlation matrix, for
example that defined by Cholesky decomposition.} in measurement $i$.
Since $C_{ij}/\sigma_i\sigma_j$ is positive definite and symmetric finding the
eigenvalues and eigenvectors is straightforward (e.g.~by Jacobi transforms).
Arrange the eigenvectors in decreasing order of eigenvalue.  The first $M$
eigenvectors in the sequence define the best $M$-dimensional subspace to use
in filtering the noise from the data.  Once increasing the dimension of the
subspace adds eigenvectors whose eigenvalues are $\ll1$ the gain for the extra
computing burden is marginal.

The advantage of optimal subspace filtering when fitting mosaiced data
is obvious.  One wishes to cover the $u-v$ plane fully with the small
effective apertures $\widetilde{A}$ obtained after summing the individual
pointings.  Thus each pointing oversamples the $u-v$ plane considerably
in terms of its larger ``unmosaiced'' $\widetilde{A}$, and these visibilities
are highly correlated.
The optimal subspace filter identifies which combinations of the visibilities
measure primarily the cosmological signal and which are mostly noise, allowing
an optimal weighting to be given to the individual visibilities sets while
retaining the processing time advantage of a smaller data set.

\section{Imaging the Microwave Sky} \label{sec:mapping}

In addition to power spectrum estimation and parameter constraints, one
goal of DASI and CBI is to image the microwave sky.
Since interferometers don't measure the temperature of the CMB sky directly,
it is necessary to ``invert'' the measured visibilities to form a sky image.
In general this inversion is not unique (the DC level for example remains
unconstrained) and so must be regularized.
On a field-by-field basis images of the sky can be made using the usual
methods of synthesis imaging in radio astronomy (e.g.~Cornwell~\cite{Cornwell})
and will form a useful data checking tool.

It is possible however to try to make a larger scale image of the microwave
sky using other (statistical) techniques.
On degree angular scales, such as probed by DASI, regularized images have
been made from the ACME/MAX (White \& Bunn~\cite{MEM}),
Saskatoon (Tegmark et al.~\cite{SaskMAP}) and CAT (Jones~\cite{Jones})
experiments by a variety of techniques.

Perhaps the most straightforward procedure for going from the visibility data
to a sky image is Wiener filtering.  Wiener filtering has been used extensively
on CMB and large-scale structure data in the past
(e.g.~Lahav et al.~\cite{LahFHSZ}, Bunn et al.~\cite{BunFHLSZ},
Zaroubi et al.~\cite{ZarHFL}, Bunn, Hoffman \& Silk~\cite{BunHofSil})
and the reader is referred to those papers for a description of the underlying
theory.  The implementation of the Wiener filter in the basis defined in
\S\ref{sec:osf} is treated in many of the references given in that section.

The Wiener filter provides an estimate of the sky temperature at a point as
a linear combination of the visibility data.
If we assume that foregrounds and point sources have been removed from the
visibility data and that the underlying sky is gaussian, then Wiener filtering
is the ``optimal'' imaging method.  Because it is linear, under the assumption
of a gaussian sky, the error matrix in the map is also easy to calculate.

To proceed, let us imagine pixelizing the sky into very many small pixels
at position ${\bf x}_\alpha$, $\alpha=1\cdots N_{\rm pix}$.  This is purely
a bookkeeping device which allows us to use a matrix notation for the Wiener
filter, the continuum limit is obtained trivially with a sum over pixels
replaced by an integral.
We denote the underlying temperature fluctuation on the sky by $t_\alpha$
which is related to the $j$th visibility by $V_j= W_{j\alpha}t_\alpha+n_j$
where the weight matrix $W_{j\alpha}$ is (see Eq.~\ref{eqn:visdef})
\begin{equation}
W_{j\alpha} = e^{2\pi i {\bf u}_j\cdot{\bf x}_\alpha} A_j({\bf x}_\alpha)
\end{equation}
and $n_j$ is the noise in the $j$th visibility.
We have labelled $A_j({\bf x})$ with a subscript $j$ to allow for the
possibility of mosaicing.

On an algorithmic note, it is computationally simpler to split the visibility
into its real and imaginary parts and modify $W_{j\alpha}$ to have cosine and
sine components so that one deals only with real variables.  The number of
data points is then doubled, and the noise covariance matrix must be modified
by a factor of 2 also.  We shall use a complex notation, with the understanding
that the implementation of the algorithm may be in terms of real valued data.

If we assume that the correlation matrices of the theory and noise are known
(or computed from the data, see Seljak~\cite{SelWF}) then the Wiener
filtered estimate of the $t_\alpha$ is
\begin{equation}
t^{WF}_\alpha = C^S_{\alpha\beta} W_{j\beta}
  \left( W_{j\gamma}C^S_{\gamma\epsilon}W_{k\epsilon} + C^N_{jk} \right)^{-1}
  V_k
\label{eqn:wfdef}
\end{equation}
where $C^S$ and $C^N$ are the signal (Eq.~\ref{eqn:cordef}) and
noise (Eq.~\ref{eqn:cndef}) correlation matrices.
The combination $W\cdot C^S \cdot W^T$ is essentially $C^V$ of
Eq.~(\ref{eqn:cvdef}), we have written it in this way to allow the possibility
of including other data sets as discussed below.
Notice that the form of the Wiener filter is ``signal/(signal+noise)'' as
is usually the case.

An expression similar to Eq.~(\ref{eqn:wfdef}) can be obtained for the
(formal) error correlation matrix of the estimates (see the above references).
The most serious problem is that the Wiener filtered image constructed in
this manner is missing large scale power.
This can be included by fitting to another data set, which retains the long
wavelength modes, at the same time as the visibility data.
In this case one extends the data vector $V_j$ to include the extra temperature
data, with the associated $W_{j\alpha}$.  In the case of a mapping experiment,
$W_{j\alpha}$ is simply the beam.  For most current degree scale experiments it
is a beam modulated on the sky by a chopping pattern.
We will defer a detailed analysis of how well we can reconstruct the sky
under various observing strategies, and the noise properties of the images,
to a future publication where we also plan to discuss the effect of sky
curvature.

\section{Polarization} \label{sec:polarization}

Neither the DASI or CBI will be polarization sensitive initially, however
some instruments operating at smaller angular scales (e.g.~the VLA) are
already sensitive to linear polarization.
The (linear) polarization of the CMB anisotropies is an important theoretical
prediction, and can encode a great deal of information about the model
(for a recent review see Hu \& White~\cite{Polar}) so we discuss it briefly
here.
Our analysis is very similar to the small-angle formalism developed in
Seljak~(\cite{SelPol}), though we caution the reader that we have a different
sign convention for the $E$- and $B$-modes, see
e.g.~\cite{OTAMM}.

We consider here only the small scale limit, so we treat the sky as a plane
with a righthanded coordinate system upon it so that the sky is the $x-y$
plane.
We define polarization in the horizontal ($\hat{x}$) and vertical ($\hat{y}$)
directions to be $Q>0$ and $Q<0$ respectively.
Polarization in the $\widehat{x}+\widehat{y}$ and $\widehat{x}-\widehat{y}$
directions is defined to be $U>0$ and $U<0$ respectively.
In terms of light traveling to us along $\hat{z}$ the intensity tensor
$I_{ij}$, with $i,j=x,y$, is
\begin{equation}
I_{ij} \propto \left\langle E_i^{*} E_j \right\rangle \propto
  T {\bf 1} + Q\sigma_3 + U\sigma_1
\end{equation}
where the angled brackets indicate an average over a time long compared with
the frequency of the wave, $E_i$ is the electric field component and
$\sigma_k$ are the Pauli matrices.  We have neglected $V\sigma_2$ since
this corresponds to circular polarization, which is not generated
cosmologically.
Under a rotation by an angle $\psi$ around $\hat{z}$ the temperature is
clearly left invariant while $Q\pm iU$ transforms as a spin-2 tensor
\begin{equation}
\left( Q\pm iU \right) \to e^{\mp 2i\psi} \left( Q\pm iU \right)
\quad .
\end{equation}
We expand $Q\pm iU$ in basis states known as spin-spherical harmonics
${}_{\pm 2}Y_{\ell m}$ with coefficients $a_{\ell m}^{\pm 2}$ in analogy with
the temperature fluctuations (Seljak \& Zaldarriaga~\cite{SelZal};
Kamionkowski, Kosowsky \& Stebbins~\cite{KamKosSte}; Hu \& White~\cite{TAMM}).
We can form states of definite parity, $a_{\ell m}^{+2}\pm a_{\ell m}^{-2}$,
which are called $E$- and $B$-mode polarization respectively
(not to be confused with the $E$ and $B$ modes of the radiation)
and which have angular power spectra $C_\ell^{E}$ and $C_\ell^{B}$ like the
temperature.  In addition the $E$-mode of the polarization is correlated with
the temperature, so there is a fourth power spectrum: $C_\ell^{TE}$.

The spin-2 spherical harmonics are better known in the context of quantum
mechanics as Wigner functions:
\begin{equation}
{}_{s}Y_{\ell m} \propto {\cal D}^{\ell}_{-s,m} =
  \left\langle \ell, -s \right| {\cal R} \left| \ell, m \right\rangle
\label{eqn:sYlmdef}
\end{equation}
where ${\cal R}=\exp[i\vec{\alpha}\cdot\vec{J}]$ is the rotation operator
and $\vec{J}$ are the angular momentum operators.  The case $s=0$ are the
well known spherical harmonics.  Thus we can relate the ${}_{\pm 2}Y_{\ell m}$
to the usual $Y_{\ell m}$ through raising and lowering operators
(Zaldarriaga \& Seljak~\cite{ZalSel}).
The raising and lowering operators are differential operators since
differentiating ${\cal R}$ with respect to its arguments inserts
$J_{\pm}\propto J_x\pm iJ_y$ inside the bracket in Eq.~(\ref{eqn:sYlmdef}).
In the flat space case the operators are trivially $\partial_x\pm i\partial_y$,
with an overall normalization $\ell^{-1}$ from the normalization of the
angular momentum states.

Going to the flat space limit (see Appendix) and acting on the Fourier
integral with $\left(\partial_x\pm i\partial_y\right)^2$ we find that
the Fourier coefficients $E_{\bf u}$ and $B_{\bf u}$ are related to the
measured $Q$ and $U$ through
\begin{equation}
\begin{array}{lcl}
-Q({\bf x}) & = &
{\displaystyle \int}
  d^2u \left( E_{\bf u}\cos 2\theta_u - B_{\bf u}\sin 2\theta_u\right)
  e^{2\pi i {\bf u}\cdot{\bf x}}\\
-U({\bf x}) & = &
{\displaystyle \int}
  d^2u \left( E_{\bf u}\sin 2\theta_u + B_{\bf u}\cos 2\theta_u\right)
  e^{2\pi i {\bf u}\cdot{\bf x}}
\end{array}
\end{equation}
where the signs reflect the definition of $E$ and $B$ in
(\cite{OTAMM}), and come from differentiating an imaginary
exponential twice.
[We caution the reader that the power spectra output by the program
{\sl CMBFAST\/} differ in the sign of $C_\ell^{TE}$ from this convention
(\cite{OTAMM}).  Obviously $C_\ell^{EE}$ and $C_\ell^{BB}$ are
invariant under the sign change.]
If we recall that changing the sign of the polarization rotates the
polarization ``vector'' through $90^\circ$ we see that the $E$-mode has
polarization always tangential in the $u-v$ plane (Seljak~\cite{SelPol};
it would be radial under a sign change).

We now have 4 non-vanishing power spectra, all diagonal in ${\bf u}$,
whose diagonal parts we can denote by $S^{TT}$, $S^{EE}$, $S^{BB}$ and
$S^{TE}$ in analogy with Eq.~(\ref{eqn:suflat}).
This is all the formalism we need to allow us to reconstruct the power
spectrum of temperature and polarization, or provide constraints on theoretical
models.  We proceed as in Hu \& White~(\cite{Polar}; see also
Zaldarriaga~\cite{ZalPOLAR}) by defining a vector of our data
$D_I=\left( T_1,Q_1,U_1;\cdots; T_N,Q_N,U_N\right)$ at each point ${\bf u}_i$.
The likelihood function for this data is analogous to
Eq.~(\ref{eqn:likelihood}) with a correlation matrix
\begin{equation}
C_{IJ}^{XY} \propto
  \int d^2u\ \widetilde{A}_i({\bf u})\widetilde{A}_j({\bf u})
  \sum_{\alpha,\beta}
  \omega^{X\alpha}(\theta_u)\omega^{Y\beta}(\theta_u)\ S^{\alpha\beta}(u)
\end{equation}
where $X,Y=(T,Q,U)$ and $\alpha,\beta=T,E,B$ with $\omega^{TT}=1$,
$\omega^{QE}=-\cos2\theta$, $\omega^{QB}=\sin2\theta$,
$\omega^{UE}=-\sin2\theta$, $\omega^{UB}=-\cos2\theta$.

Given a set of measurements one can reconstruct the power spectrum of the
temperature or the polarization as described in \S\ref{sec:quadratic}.
Alternatively one can provide upper limits (or measurements) of a bandpower
for polarization or temperature by calculating ${\cal L}(Q^2)$ where
$Q^2$ is the amplitude of a flat power spectrum
(i.e.~$\ell(\ell+1)C_\ell=$constant) for the $TT$, $TE$, $EE$ and $BB$ power
spectra.
[If information on the temperature is absent one merely drops the $X=T$ entries
of the correlation matrix above.]
If one believes that there is no foreground contamination, the cosmological
signal on small scales should be dominated by $E$-mode polarization and so
one can set $Q_B=0$ in evaluating the likelihood.

\section{Discussion} \label{sec:discussion}

The theoretical study of CMB anisotropies has advanced considerably in
recent years, both in terms of theoretical predictions and comparison with
observational data.
In this paper we have developed some of the formalism necessary for the
analysis of interferometer data, such as will be returned by VSA, DASI, CBI
and VLA, in the modern language of anisotropy (temperature and polarization)
power spectra.
Although the fundamental entities measured by an interferometer differ
considerably from those measured in single-dish experiments, much of the
analysis can be presented in an analogous manner including power spectra,
window functions and the like.  This allows one to use directly the
sophisticated analysis techniques which already exist for single-dish
experiments.

The formalism presented here is based upon the ``flat sky'' approximation
and is thus best suited to the study of small-scale anisotropies.  In this
regime interferometers can provide high-sensitivity measurements of the
high-$\ell$ peaks in the angular power spectrum, effects of gravitational
lensing, the damping tail of the anisotropies and proto-galaxy formation and
second order anisotropies.  For CBI the cosmological signal would be higher
if the universe had significant spatial curvature since this would shift the
features in the angular power spectra to smaller angular scales, however for
both VSA and DASI significant signal is expected simply based on existing
measurements.

Several outstanding problems remain, all associated with extending the
framework of this paper to larger angular scales where the curvature of the
sky and the full 3D nature of the Fourier Transforms become important.
We intend to return to these issues in a future paper.

\bigskip
\acknowledgments  
We would like to acknowledge useful conversations with Wayne Hu,
Douglas Scott and Max Tegmark.  We thank Lyman Page for many comments which
helped to clarify the paper.
M.W.~thanks Pedro Ferreira for conversations during the initial phases
of this work.

\appendix

\section{Flat Sky Approximation}

The derivation of the correlation function Eq.~(\ref{eqn:suflat}) given
in the text is sufficient for the purposes of this paper.  However,
sometimes it is convenient to treat the sky as flat and replace spherical
harmonic sums with Fourier Transforms at the temperature (rather than the
2-point function) level.
We give some details of this in this Appendix, the reader is also referred
to Zaldarriaga \& Seljak~(\cite{ZalSel}).

In the flat sky approximation we replace the spherical polar coordinates
$(\theta,\phi)$ by radial coordinates on a plane:
$r\equiv 2\sin\theta/2\approx\theta$ and $\phi$.  We can exchange our
indices $(\ell,m)$ for a 2D vector ${\bf l}$ with length $\ell$ and
azimuthal angle $\varphi_\ell$ and define
\begin{equation}
a({\bf l}) \equiv \sqrt{ 4\pi\over 2\ell+1 } \sum_m i^{-m} a_{\ell m}
  \ e^{im\varphi_\ell}
\end{equation}
with $a(-{\bf l})=a^{*}({\bf l})$.
We now expand our temperature field in terms of multipole moments as usual
with
\begin{equation}
\sum_{\ell m} a_{\ell m} Y_{\ell m} = {1\over 2\pi} \int d\ell\, d\varphi_\ell
\ a({\bf l}) \sqrt{2\ell+1\over 4\pi} \sum_m i^m e^{-im\varphi_\ell} Y_{\ell m}
\end{equation}
where we have replaced the sum over $\ell$ with an integal.

Writing the sum over $m$ in terms of $m<0$ and replacing the associated
Legendre polynomials with Bessel functions using Gradshteyn \& Ryzhik
(\cite{GraRyz}; Eq.~8722(2)) we have
\begin{equation}
\sum_{\ell m} a_{\ell m} Y_{\ell m} \approx
{1\over (2\pi)^2} \int \ell d\ell \, d\varphi_\ell
\ a({\bf l}) \left[
  J_0(\ell\theta) + 2\sum_{m=1}^\infty
  i^m J_m(\ell\theta)\cos m(\phi-\varphi_\ell)
\right] \quad .
\end{equation}
The term in square brackets is simply the Rayliegh expansion of a plane
wave so we finally obtain, writing ${\bf l}=2\pi {\bf u}$,
\begin{equation}
{\Delta T\over T}({\bf x}) = \sum_{\ell m} a_{\ell m} Y_{\ell m} \approx
\int d^2u \ a({\bf u}) \exp\left[ 2\pi i{\bf u}\cdot{\bf x} \right]
\label{eqn:flatsky}
\end{equation}
and of course the two point function for $a({\bf u})$ is diagonal:
\begin{equation}
\left\langle a^{*}({\bf u}) a({\bf w}) \right\rangle
= C_\ell \delta^{(2)}({\bf u}-{\bf w}) \quad .
\end{equation}

These expressions can then be used to derive Eq.~(\ref{eqn:suapprox}), though
the derivation presented above Eq.~(\ref{eqn:suflat}) uses more controlled
approximations.
The generalization of Eq.~(\ref{eqn:flatsky}) to polarization is presented
in \S\ref{sec:polarization}.

\end{document}